\documentclass[USenglish]{article}

\usepackage[%
english,    %
]{babel}

\usepackage[%
utf8        
]{inputenc}

\usepackage[%
T1          
]{fontenc}

\usepackage{lmodern}            

\usepackage{textcomp}

\usepackage{microtype}

\usepackage[                    %
affil-it                        
]{authblk}

\usepackage[%
format=plain,           
labelsep=colon,               
justification=justified,      
singlelinecheck=true,         
labelfont=bf                  
]{caption}

\usepackage{subcaption}

\usepackage[%
  style=authoryear,             
  natbib=true,                  
  safeinputenc=true,            
  maxbibnames=20,               
  minbibnames=10,               
  maxcitenames=2,               
  hyperref=true,                
  backref=false,                
  isbn=false,                   
  url=false,                    
  doi=false,                    
  eprint=false,                 
  firstinits=true,              
  uniquename=false,             
  uniquelist=false,             
  sorting=nyt                   
]{biblatex}

\bibliography{references.bib}        

\DeclareNameAlias{sortname}{last-first} 

\usepackage{amsmath}
\usepackage{amsthm}
\usepackage{amsfonts}
\usepackage{amssymb}
\usepackage{amstext}

\usepackage{mathtools}

\usepackage{physics}

\usepackage{multirow}

\renewcommand{\epsilon}{\varepsilon}

\usepackage{siunitx}
\begin{document}

\microtypesetup{protrusion=false}
\title{Intercomparison of Warm-Rain Bulk Microphysics Schemes using Asymptotics}

\author[1]{Juliane Rosemeier}
\author[2]{Manuel Baumgartner}
\author[1]{Peter Spichtinger}

\affil[1]{                      %
  Institute for Atmospheric Physics, Johannes Gutenberg University, Mainz, Germany
}
\affil[2]{                      %
  Data Center, Johannes Gutenberg University, Mainz, Germany
}

\date{\today}
\maketitle
\microtypesetup{protrusion=true}

\begin{abstract}
  Clouds are important components of the atmosphere. Since it is
  usually not possible to treat them as ensembles of huge numbers of
  particles, parameterizations on the basis of averaged quantities (mass
  and/or number concentration) must be derived. Since no
  first-principles derivations of such averaged schemes are available
  today, many alternative approximating schemes of cloud processes
  exist. Most of these come in the form of nonlinear differential
  equations. It is unclear whether these different cloud schemes behave
  similarly under controlled local conditions, and much less so when
  they are embedded dynamically in a full atmospheric flow model. We use
  mathematical methods from the theory of dynamical systems and
  asymptotic analysis to compare two operational cloud schemes and one
  research scheme qualitatively in a simplified context in which the
  moist dynamics is reduced to a system of ODEs. It turns out that these
  schemes behave qualitatively differently on shorter time scales,
  whereas at least their long time behavior is similar under certain
  conditions. These results show that the quality of computational
  forecasts of moist atmospheric flows will generally depend strongly on
  the formulation of the cloud schemes used.
\end{abstract}


\section{Introduction}
\label{sec:intro}

Clouds constitute one of the most important but rather unknown
components of the Earth-Atmosphere system. They influence the
hydrological cycle and also the energy budget of the system due to
interaction with radiation. Clouds consist of a huge number of water
particles of different phases (liquid and/or solid), therefore the
simulation of the trajectories for each individual water particle requires too
much computational effort and statistical approaches must be used in
order to treat the system in an appropriate way. One common approach
is the use of a size or mass distribution of cloud particles. However,
to date there is no closed description of the time evolution of a cloud size
distribution available; especially there is no commonly accepted or
even formulated description of the sources and sink terms. Usually
only collisional terms are treated in such approaches
\citep[e.g.,][]{beheng2010}. Since the treatment of size distributions
is also numerically difficult and expensive, one uses averaged
quantities like mass and number concentrations as cloud
variables. In terms of an underlying size distribution, these
variables are just general moments of the distribution. The treatment
of collisional source and sink terms for averaged variables cannot be
described uniquely. Since there is no systematic, closed
derivation for cloud schemes available at the moment,
i.e. based on first principles
which uses averaged (also called ``bulk'') variables, many
different formulations for cloud processes are found in literature,
differing in the description of the basic processes,
especially collision processes.  For the use in operational numerical
weather prediction (NWP) models, simple cloud schemes are
implemented. Most NWP models use so called single moment schemes,
i.e. cloud schemes with mass concentrations as averaged variables.
Due to phase transition processes, there is latent heat release in
clouds, which drives buoyancy and therefore affects the
atmospheric motions. On the other hand, the mass of cloud
condensate modifies the cloud buoyancy to smaller values. Thus, the
predictability of moist atmospheric flows may be crucially affected by
clouds and, more important, on the representation of clouds in the
models. The representation of cloud processes affects the macroscopic
structure of clouds; due to diabatic heating (radiative feedback and
latent heating) in inhomogeneous clouds this might also affect motions
on larger scales. Therefore we can assume that the formulation of
cloud processes can also affect atmospheric processes on larger
scales, changing the predictability of clouds, precipitation and
atmospheric motions.


In this study, we investigate different cloud schemes (two operational
schemes and one research scheme) about their qualitative behavior
under very idealized conditions. These schemes are written as systems
of ordinary differential equations (ODEs). Including rain
sedimentation results in partial differential equations (PDEs), unless
sedimentation is represented in a simplified way. This is the case in
the present study and will be explained later.  The main purpose is to
identify possible equilibrium states and other qualitative properties
of these schemes in the sense of dynamical systems. Since the models
are coupled to atmospheric flows via PDEs, their behavior may impact
flow characteristics. Atmospheric flows are represented by the
Navier-Stokes equations or some valid approximations thereof. It is a priori
not clear how large the differences may be when two cloud schemes with
different qualitative behavior are used. The impact of qualitatively
different formulations for cloud schemes on the final flow cannot be
derived easily. It is quite obvious that processes represented by
Laplacian operators might lead to stabilisation of unstable
equilibria. On the other hand, the seminal work by \citet{turing1952}
showed that diffusion terms (i.e. Laplacian operators) might also lead
to instabilities, and in turn to pattern formation \citep[cf.,
e.g.,][]{cross_hohenberg1993}. However, the determination of
equilibrium states of underlying dynamical systems is necessary for
such further investigations, which are beyond the scope of our study.




We investigate single moment bulk schemes for warm clouds, i.e. for clouds
containing only water in liquid phase. The equations of the schemes
describe the evolution of the averaged mass. The mass concentration or
mixing ratio of a species $x\in\{v,c,r\}$ is defined by
$q_x\coloneqq\frac{M_x}{M_a}$, where $M_x$ denotes the mass of species
$x$ and $M_a$ denotes the mass of dry air. Vapor is indicated by the
index $v$, while the index $c$ denotes cloud droplets and index $r$
indicates rain drops, respectively. Since collision processes must be
formulated via averaged quantities, traditionally these schemes
discriminate between small, non-sedimenting cloud particles and
large rain drops, which fall out due to gravitational acceleration.  Such
one-moment cloud schemes are found in the operational forecast models
IFS, run by the European Center for Medium Range Weather Forecasts
(ECMWF), and COSMO, run by the German
Weather Service (DWD), and are largely inspired by the early work from
\citet{kessler1969}, who already made the distinction between cloud
droplets and rain drops.

In the sequel, we consider the following standard description of a
one moment bulk scheme for a warm cloud in a zero-dimensional parcel
framework
\begin{subequations}
  \label{eq:intro_formal_governing_system}
  \begin{align}
    \dv{q_c}{t} &= C - A_1 - A_2,\label{eq:intro_formal_governing_system-qc}\\ 
    \dv{q_r}{t} &=  \qquad A_1 + A_2 - E + B-D,\label{eq:intro_formal_governing_system-qr}
  \end{align}
\end{subequations}
containing the following cloud processes:
\begin{itemize}
\item Condensation $C$: Growth of cloud droplets by diffusion of water vapor,
\item Autoconversion $A_1$: Collision of cloud droplets which coalesce and ultimately form large rain drops
\item Accretion $A_2$: Collection of cloud droplets by a falling rain drop,
\item Evaporation $E$: rain drops grow or shrink due to phase transitions,
\item Rain flux from above $B$: Rain falling from above into the air parcel
  under consideration,
\item Sedimentation of rain $D$: Rain falls out of the air parcel.
\end{itemize}
We point out here that $B$ and $D$ appear separately. We will prescribe $B$,
being an ad-hoc assumption. This is necessary because $B$
describes the impact of the control volumes higher up. If the flux $B$
were to be expressed in terms of dynamical variables, we would couple
neighboring cells and end up with a PDE. However in this study, we
focus on the local evolution. Finally, $B$ can be considered
as an external forcing. The term $D$ is tied to the
local conditions in the control volume. This explains why $D$ is
split from $B$.  

Note that our description of diffusional processes includes always
two scenarios, i.e. supersaturation (growth of water droplets) and
subsaturation (shrinking of water droplets). We refer to the
diffusion process for cloud droplets as condensation (since here the
supersaturation regime is relevant), whereas evaporation denotes the
diffusion process for rain drops, which is more important for
rain. Of course, rain drops can also grow by diffusion (and this
takes place in our scenarios if $S>0$); however, this process leads
to very small changes in water phases and can usually be neglected,
as we will see in the asymptotic analysis.

Inspecting the various cloud schemes in the literature reveals, that
most formulations of cloud schemes contain descriptions of those
processes, but their mathematical description differs. Only the
process of condensation has an accepted derivation from Maxwellian
growth theory \citep[][]{maxwell1890,book_rogers_yau1989}, although it
is exactly this process which is circumvented to include in
operational forecast models, because it imposes a severe time step
restriction. In operational models, one usually replaces
condensation by so-called saturation
adjustment
\citep[e.g.,][]{mcdonald1963, asai1965, langlois1973,soong_ogura1973, yau_austin1979, rutledge_hobbs1983,kogan_martin1994, bryan_fritsch2002},
ensuring saturated conditions or conditions, being as close to
saturation as possible by condensing or evaporating cloud droplet mass.
In our study, we will include this process explicitly.  Apart from the
operationally employed cloud schemes within the IFS
\citep{ifs_doc_physical_parameterization} and COSMO
\citep{cosmo_doc_physical_parameterization} models, we consider a
cloud scheme introduced in the study of \citet{wacker1992}. In her
study, Wacker analyzed a cloud scheme with regard to its equilibrium
behavior. This was the first study into this direction,
motivating us to include her model in our study, since we also analyzed
the other two models in this respect.

The goal of our study is to analyze the three models
and understand their characteristic behavior of representing
warm clouds. We do not intend to rate the models, but investigate
their characteristic response with respect to different regimes.
This helps to interpret the results of a simulation, carried out with
the IFS or the COSMO model regarding warm clouds. Since the models
lead to different equilibrium states, this must be taken into account,
if more complex model simulations are investigated and compared. If
the underlying cloud schemes do not agree qualitatively and/or
quantitatively, one cannot assume that the atmospheric models,
which introduce couplings between their underlying PDEs and the
cloud schemes, will produce similar results.

This study is organized as follows.
In section \ref{sec:models} we describe a generic cloud scheme,
which contains all three cloud schemes as special cases. Section
\ref{sec:qualitative} is dedicated to the analysis of the qualitative
behavior of the cloud schemes. As will become clear, the qualitative
behavior is linked to the long time behavior. Using asymptotic
techniques in section \ref{sec:asymptotic}, we derive reduced equations
for several regimes describing the dominant behavior on selected
time scales for relevant environmental conditions. 
A discussion of these results is found in section \ref{subsec:discussion}
and a more general conclusion in section \ref{sec:conclusion}.

\section{The Cloud Schemes}
\label{sec:models}

In this study, we consider the three one moment cloud schemes for warm cloud
microphysics, found in \citet{wacker1992} (in the following referred to as
``Wacker''), the scheme used in the IFS model (referred to as ``IFS'') and
the scheme incorporated in the COSMO model (referred to as ``COSMO'').
In section \ref{subsec:generic}, we present a generic cloud scheme, containing
all the before mentioned cloud models as a special case. Section
\ref{subsec:nondimensionalization} contains the description of a
nondimensionalization of the generic cloud scheme. The convenience of
nondimensionalization is explained later. Section
\ref{subsec:specialized}  refers to the specialized cloud
schemes in this context.

\subsection{A Generic Cloud Scheme}
\label{subsec:generic}

To describe our generic cloud scheme, we have to model all processes
from equation \eqref{eq:intro_formal_governing_system}. We denote the
supersaturation with respect to water by
$S=\frac{e}{e_{\mathrm{sat}}}-1$, where $e$ is the partial pressure of
water vapor and 
$e_{\mathrm{sat}}$ is the saturation vapor pressure over a flat 
surface of water.  For the condensation rate $C$ in equation
\eqref{eq:intro_formal_governing_system} we set
\begin{equation}
  \label{eq:generic_def_C}
  C\coloneqq cSq_c 
\end{equation}
with a suitable constant $c$, as suggested by \citet{wacker1992};
see appendix \ref{sec:appendix_condensation}.
Note that we use this term in order to replace the saturation adjustment
scheme, used in the operational forecast models.  
Supersaturation inside clouds can be produced by permanent cooling of
the system (e.g. by a vertical updraft), which is not completely
balanced by diffusional growth. Low vertical updrafts (e.g. along warm
fronts) produce tiny supersaturations, whereas high vertical velocities
(e.g. in warm conveyor belts or convective systems) lead to quite
substantial supersaturation. The supersaturation can be maintained to
be almost constant over a certain timescale, as can be found in theoretical
studies \citep{korolev_mazin2003} or from box model simulations
(K. Diehl, pers. comm.). However,  the timescale depends crucially on the
strength of the updraft velocity.


In our study, we assume constant supersaturation. This assumption may
be violated in applications, however this study should be regarded as
a consistency test, because consistency is exactly (and actually the
only thing) what models, and in particular cloud schemes, can
accomplish \citep{oreskes_etal94}.
In our context, consistency means that cloud schemes which are designed
to represent the same physical processes and are similarly formulated,
lead to the same or at least similar results and qualitative behavior.
As we will see in our analysis, the cloud schemes can produce rather
different results despite their quite similar formulation
(they are all special cases of the generic formulation in equation
\eqref{eq:generic_generic_model}). In order to study the consistency,
assuming constant supersaturation seems reasonable. While constant
supersaturation is assumed for indefinite time 
in the first analysis, we will later restrict the analysis (and thus
the validity of this assumption) to distinct timescales.

Autoconversion depends on cloud water mixing ratio only as the rain
already present in the volume should not affect
autoconversion. However, there could be a threshold cloud water as in
the traditional Kessler formulation.  The autoconversion $A_1$ in
equation \eqref{eq:intro_formal_governing_system} may be modeled as
\begin{equation}
  \label{eq:generic_def_A1}
  A_1\coloneqq a_1q_c^{\gamma}.
\end{equation}
In contrast, accretion implies collisions between cloud water and rain, and thus
its representation should depend on both mixing ratios.
We define $A_2$ as
\begin{equation}
  \label{eq:generic_def_A2}
  A_2\coloneqq a_2q_c^{\beta_c}q_r^{\beta_r},
\end{equation}
where $a_1,\,a_2$ are positive coefficients and $\gamma,\,\beta_c,\,\beta_r$ are
constant positive exponents.
An analogy of
these ideas is employed in modeling predator-prey population dynamics
\citep[cf., e.g.,][]{murray2002}.

The process of evaporation $E$, which also includes diffusional growth
of large rain drops for $S>0$,
is modeled similar to the condensation
process as
\begin{equation}
  \label{eq:generic_def_E}
  E\coloneqq -\qty(e_1q_r^{\delta_1}+e_2q_r^{\delta_2})S
\end{equation}
with coefficients $e_1,\,e_2$ and constant positive
exponents $\delta_1,\,\delta_2$. Modeling evaporation as a sum of two
individual terms results from taking ventilation into account
\citep[see e.g.][]{seifert_beheng2006a}: the description of diffusional
growth usually assumes a calm environment about the drop. This assumption
is relaxed by introducing the ventilation coefficient 
in the formulation $f_v=a_v+b_v(p,T,q_r)$,
capturing the influence of air motion on diffusional growth
\citep[][]{book_pruppacher_klett2010,pruppacher_rasmussen1979}.
Since the ventilation
coefficient is multiplied to the growth equation, this yields a
generic sum as in \eqref{eq:generic_def_E}.

We model sedimentation $D$ of rain drops by
\begin{equation}
  \label{eq:generic_def_D}
  D\coloneqq dq_r^{\zeta}
\end{equation}
with a coefficient $d$ and a constant positive exponent $\zeta$. The
coefficient $d$ is given as $d = \frac{v}{h}$, where 
$v$ is the model specific parameterization of the terminal fall
velocity for rain drops and $h$ is the height of the considered
control volume or air parcel,
assumed as $h = \SI{1000}{\meter}$ for all cloud schemes
in this study.
Moreover, we assume that $B$, the rain sedimentation flow from above, is
constant.


Substituting the definitions
\eqref{eq:generic_def_C},
\eqref{eq:generic_def_A1},
\eqref{eq:generic_def_A2},
\eqref{eq:generic_def_E},
\eqref{eq:generic_def_D}
for the individual processes into equation
\eqref{eq:intro_formal_governing_system} yields the generic cloud
scheme
\begin{subequations}
  \label{eq:generic_generic_model}
  \begin{align}
    \dv{q_c}{t} &= cSq_c - a_1q_c^{\gamma} - a_2q_c^{\beta_c}q_r^{\beta_r},\label{eq:generic_generic_model-qc}\\ 
    \dv{q_r}{t} &=  \qquad a_1q_c^{\gamma} + a_2q_c^{\beta_c}q_r^{\beta_r} +\qty(e_1q_r^{\delta_1}+e_2q_r^{\delta_2})S  + B-dq_r^{\zeta}.\label{eq:generic_generic_model-qr}
  \end{align}
\end{subequations}
The coefficients in the generic scheme \eqref{eq:generic_generic_model}
usually depend on environmental pressure and temperature whereas all
exponents are fixed constants. As a consequence of our generic formulation,
every choice of positive coefficients and exponents
yields a possible cloud scheme although not every choice also represents
a physically meaningful cloud scheme.
The three cloud schemes of our study fit into this framework, as
becomes clear in section \ref{subsec:specialized}.
The term $B$, indicating the rate of rainfall from above into the current
air parcel, is not specified by the scheme, but determined
by the conditions within the air parcel or grid box above.

\subsection{Nondimensionalization}
\label{subsec:nondimensionalization}

For a
rigorous mathematical analysis of the equations, nondimensionalization
is a common tool. It assures that the variables are normalized to the
same order of magnitude; in addition, dominant processes can be
identified. The nondimensionalization might lead to several small
parameters in the equations, which are a key issue for the asymptotic
analysis, see section~\ref{sec:asymptotic}.

Here, we describe the nondimensionalization of the generic cloud
scheme \eqref{eq:generic_generic_model} using a reference value
$t_{\mathrm{ref}}=\SI{1}{\second}$ for time and
$q_{\mathrm{ref}}=\SI{e-4}{\kilogram\per\kilogram}$
for the mixing-ratios. In the sequel, we indicate a quantity
with a prime, if this quantity has a physical dimension or is
unscaled. So, when $\Psi$ is a quantity which is scaled or has a
physical dimension, then $\Psi'$ is the corresponding unscaled
nondimensional quantity. The only exception are
the reference quantities $t_{\mathrm{ref}},\,q_{\mathrm{ref}}$. The corresponding
nondimensional quantity $\Psi$ is defined by
$\Psi\coloneqq\frac{\Psi'}{\Psi_{\mathrm{ref}}}$.
Using the chain rule, we arrive at the time derivative
$\dv{q_c}{t}\qty(t)=\dv{t}\qty(\frac{q_c'(t\cdot t_{\mathrm{ref}})}{q_{\mathrm{ref}}})=\frac{t_{\mathrm{ref}}}{q_{\mathrm{ref}}}\dv{q_c'}{t'}\qty(t')$
for the cloud droplet mixing-ratio.
The derivative of $q_r$ is derived in the same fashion. As a result, we obtain
the same system as in equation \eqref{eq:generic_generic_model}, but with the
nondimensional coefficients
\begin{equation}
  \label{eq:nondimensionalization_coeff}
  \begin{aligned}
    c&=t_{\mathrm{ref}}c',& a_1&=t_{\mathrm{ref}}q_{\mathrm{ref}}^{\gamma-1}a_1',& a_2&=t_{\mathrm{ref}}q_{\mathrm{ref}}^{\beta_c+\beta_r-1}a_2',\\
    e_1&=t_{\mathrm{ref}}q_{\mathrm{ref}}^{\delta_1-1}e_1',& e_2&=t_{\mathrm{ref}}q_{\mathrm{ref}}^{\delta_2-1}e_2',& d&=t_{\mathrm{ref}}q_{\mathrm{ref}}^{\zeta-1}d'
  \end{aligned}
\end{equation}
and $B=\frac{t_{\mathrm{ref}}}{q_{\mathrm{ref}}}B'$.
This normalization leads to variables $q_c,\,q_r$ comparable to $1$;
thus, the terms on the right hand side can be compared quantitatively.

\subsection{The Specialized Cloud Schemes}
\label{subsec:specialized}

As explained in a preceding section, every choice of the coefficients and
exponents yields a cloud scheme and, in particular, the scheme by Wacker,
the COSMO
and IFS schemes are special cases of the generic cloud scheme
\eqref{eq:generic_generic_model}. We remark, that we employ the same
condensation term $C$ for all schemes. Note that the operational
schemes do not use an explicit parameterization for the condensation,
since a saturation adjustment technique is applied. In the original
work from \citet{wacker1992}, the coefficient $c$ is set constant and
does not depend on the environmental conditions.

\begin{figure}
  \centering
  \includegraphics[width=0.8\textwidth]{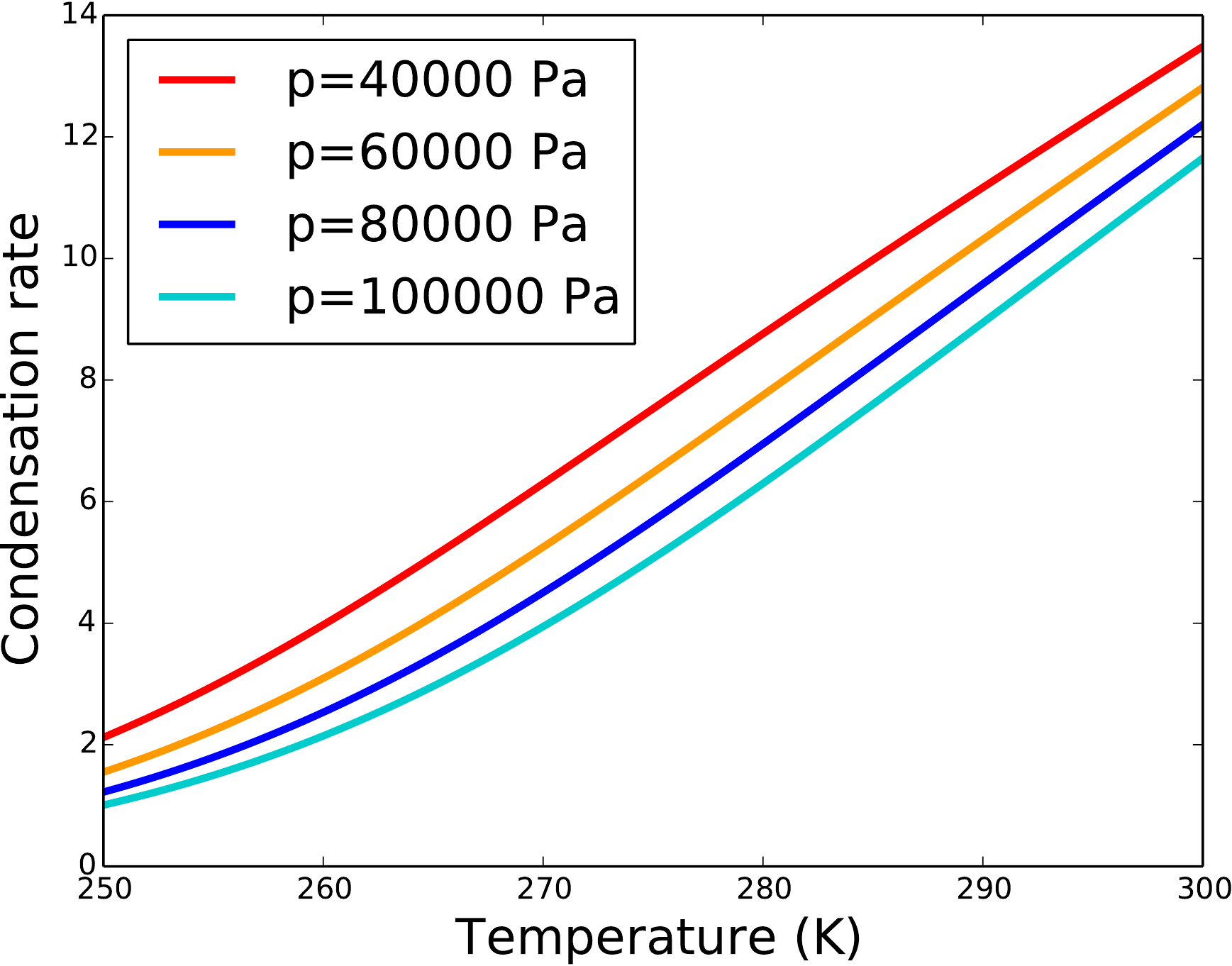}
  \caption{Nondimensional condensation rate $c$ as a function of
    temperature for various values of pressure $p$.}
  \label{fig:specialized_condensation_rate}
\end{figure}

\begin{table}
  \centering
  \begin{tabular}{c|c|c}
    Scheme & Non-constant & Constant \\
    \hline 
    Wacker & $c$ & $a_1,\,a_2,\,e_1,\,e_2,\,d$ \\
    Cosmo & $c,\,a_2,\,e_1,\,e_2,\,d$ & $a_1$ \\
    IFS & $c,\,e_1,\,e_2$ & $a_1,\,a_2,\,d$\\
  \end{tabular}
  \caption{Summary of the constant and non-constant coefficients for the three cloud schemes. A coefficient is labelled ``non-constant'', if there is a dependency on environmental temperature or pressure.}
  \label{tab:specialized_const_and_nonconst}
\end{table}

Figure
\ref{fig:specialized_condensation_rate}
shows the dependence of the nondimensional condensation rate $c$ on
environmental 
pressure and temperature. It is obvious that the condensation rate depends
strongly on temperature, motivating the consideration of several regimes.
Table
\ref{tab:specialized_const_and_nonconst}
summarizes the constant and non-constant coefficients for the three cloud
schemes considered in this study. Appendix \ref{sec:appendix_coeff}
summarizes the values of the constant coefficients and the exponents and
illustrates the dependency of the non-constant coefficients on the
environmental conditions. It should be emphasized that all non-constant
coefficients, apart from the condensation rate, depend only very weakly on 
environmental conditions.
In addition, the Wacker scheme neglects the evaporation process
for rain drops, i.e. for this
scheme we have $e_1=e_2=0$.

\section{Qualitative Behavior}
\label{sec:qualitative}

Without external forcings, a non-precipitating
warm cloud may be thought of as being in
thermodynamical equilibrium \citep{book_pruppacher_klett2010},
implying a short transient time.
If external forcings (e.g. vertical upward motions inducing a source
for supersaturation) and external sources and sinks, like sedimentation
of rain, are included, the situation will change. However, if the
forcing on the system is
constant, we may again expect the system to evolve into an equilibrium
state within some time frame, although this new equilibrium state
is different from the thermodynamic equilibrium.
Therefore we can anticipate that the
equilibrium states of the cloud scheme 
\eqref{eq:generic_generic_model}
 represent a good approximation of the description of a warm cloud.
In this case, an equilibrium state $(q_{c,e},\,q_{r,e})$ is defined
by the requirement $F(q_{c,e},\,q_{r,e})=(0,\,0)$, provided
$F$ denotes the right-hand side of the ordinary differential equation
\eqref{eq:generic_generic_model}. Geometrically, an equilibrium state
is a point (or even a manifold) in the phase space, where the
corresponding solution of the differential equation is constant, if
the value of the equilibrium state is the given initial condition.

Linearization around an equilibrium state is a common method to
determine the quality of the equilibrium state. When the equilibrium
state is a single point $(q_{c,e},\,q_{r,e})$, the characteristics of
this point are given by the eigenvalues
$\lambda_1,\,\lambda_2\in\mathbb C$ of the derivative
$(DF)_{(q_{c,e},\,q_{r,e})}$. The classification of two dimensional
systems is straightforward \citep{hirsch_etal2013}. In the case of two
dimensions, either both eigenvalues are real
$\lambda_{1,2}\in\mathbb{R}$ or complex $\lambda_{1,2}\in\mathbb{C}$;
in the latter case, $\lambda_2$ is the complex conjugate of
$\lambda_1$.  The equilibrium point is only stable if the real parts
of both eigenvalues are negative. If at least one eigenvalue admits a
positive real part, the equilibrium point is unstable. In our
analysis, we determine equilibrium points and compute the Jacobian at
the equilibrium point as well as the eigenvalues of the Jacobian. It
turns out that in certain cases the eigenvalues have non-vanishing
imaginary part, implying solution trajectories in the 
$q_c,q_r$-phase space that spiral around the equilibrium point.
The frequency of this spiralling
motion can be obtained by the imaginary part of the two complex
conjugate eigenvalues.

Instead of equilibrium points also limit cycles
(i.e. one dimensional closed curves) may occur, leading to
oscillating behavior of solution trajectories of the system.
Note that the quality
of an equilibrium point may change by changing the values of the
constants and exponents of the cloud scheme. The investigation
of all such changes involves a complete bifurcation analysis and
is beyond the scope of this study.

The derivative $(DF)_{\qty(q_c,\,q_r)}$ of the generic system
\eqref{eq:generic_generic_model} at the point
$\qty(q_c,\,q_r)$ is given by
\begin{equation}
  \label{eq:qualitative_jacobi}
  \begin{pmatrix}
    Sc-a_1\gamma q_c^{\gamma-1}-a_2\beta_cq_c^{\beta_c-1}q_r^{\beta_r} & -a_2\beta_rq_c^{\beta_c}q_r^{\beta_r-1} \\
    a_1\gamma q_c^{\gamma-1}+a_2\beta_cq_c^{\beta_c-1}q_r^{\beta_r} & a_2\beta_rq_c^{\beta_c}q_r^{\beta_r-1}+H(q_c,q_r)-d\zeta q_r^{\zeta-1}
  \end{pmatrix}
\end{equation}
with
\begin{equation}
  \label{eq:qualitative_abbr_der_evap}
  H(q_c,\,q_r)\coloneqq \qty(e_1\delta_{1}q_r^{\delta_{1}-1}+e_2\delta_{2}q_r^{\delta_{2}-1})S.
\end{equation}


In this section, we analyze the equilibrium points of the three
cloud schemes. For our analysis, we fix the supersaturation and
environmental conditions.

For the Wacker scheme, all equilibrium points can be computed
analytically
\citep[see][]{wacker1992},
whereas the other schemes use rational exponents,
leading to polynomial equations of typically high order, which
may admit many solutions. In section
\ref{subsec:qualitative_trivial} we investigate a trivial
equilibrium point and in section
\ref{subsec:qualitative_nontrivial} we discuss a non-trivial
equilibrium point. For the examination of the non-trivial
equilibrium point, we make use of numerical approximations.

Although the qualitative analysis in this section only reveals 
the behavior of the cloud schemes in the limit $t\to\infty$,
it nevertheless gives insight about the behavior of the schemes.
In the next section, we address the behavior on shorter
timescales.

\subsection{The trivial Equilibrium Point}
\label{subsec:qualitative_trivial}

Inspecting equation \eqref{eq:generic_generic_model-qc}, we observe that this
equation is trivially satisfied when $q_{c,e}=0$. In this case, there
are no cloud droplets, resembling the cloud-free case.
It follows from \eqref{eq:generic_generic_model-qr} that the corresponding value $q_{r,e}$
satisfies the nonlinear equation
\begin{equation}
  \label{eq:qualitative_trivial_eq_qr}
  0=\qty(e_1q_{r,e}^{\delta_1}+e_2q_{r,e}^{\delta_2})S+B-dq_{r,e}^{\zeta}.
\end{equation}

We neglect the evaporation process in
\eqref{eq:qualitative_trivial_eq_qr} in the discussion of the
equilibrium points (i.e. we assume $e_1=e_2=0$ and $H=0$). 
It will become clear in section \ref{sec:asymptotic} that the neglect
of the evaporation process in our three cloud schemes is a reasonable
assumption, since its impact is only minor. 

Omitting the evaporation process in \eqref{eq:qualitative_trivial_eq_qr},
we arrive at the analytical solution
\begin{equation}
  \label{eq:trivial_fp}
  q_{r,e}=\qty(\frac{B}{d})^{\frac{1}{\zeta}}.
\end{equation}
This trivial equilibrium
point has the following physical interpretation: rain falls into the air
parcel from above, cannot interact with cloud droplets since no
cloud droplets are present, and falls out of the parcel.

As shown in table \ref{tab:appendix_coeff}, the exponents
$\gamma$ and $\beta_c$ of the specialized cloud schemes satisfy
$1\leq\gamma,\,\beta_c$. 
Consequently, we are prompted to assume
$1\leq\gamma,\,\beta_c$ in the following. 
However, we remark that the COSMO scheme satisfies $\beta_r<1$. One may
also argue that the condensation term $cSq_c$ could be replaced by
$cSq_c^{\frac{1}{3}}$, see appendix \ref{sec:appendix_condensation}.
Such choices are possible, but destroy the Lipschitz continuity
of $F$ for $q_c=0$, implying the possible existence of more than one exact
solution. A more detailed analysis of this topic is beyond the scope
of this paper, so we stick with the assumption $1\leq\gamma,\,\beta_c$.

Assuming $\gamma,\,\beta_c>1$, omitting the $H$-term and
substituting $q_{c,e}=0$, the derivative 
\eqref{eq:qualitative_jacobi} yields the matrix
\begin{equation}
  \label{eq:trivial_DF}
  (DF)_{\qty(0,\,q_{r,e})}=\begin{pmatrix}cS & 0 \\ 0 & -d\zeta q_r^{\zeta-1}\end{pmatrix},
\end{equation}
implying that the trivial equilibrium point is unstable
(note: $S>0$). This observation
applies to the IFS scheme, see table \ref{tab:appendix_coeff}.

If $\gamma=1$ or $\beta_c=1$, being true for the Wacker and the
COSMO scheme, the derivative is given by the matrix
\begin{equation}
  \label{eq:trivial_DF_specialized}
  (DF)_{\qty(0,\,q_{r,e})}=\begin{pmatrix}\lambda_{1} & 0 \\ \eta & -d\zeta q_{r,e}^{\zeta-1}\end{pmatrix}
\end{equation}
with
\begin{itemize}
\item $\lambda_{1}=Sc-a_1$ and $\eta=a_1$ for the case $\gamma=1,\,\beta_c>1$,
\item $\lambda_{1}=Sc-a_2\beta_cq_{r,e}^{\beta_r}$ and $\eta=a_2q_{r,e}^{\beta_r}$ for the case $\gamma>1,\,\beta_c=1$,
\item $\lambda_{1}=Sc-a_1-a_2q_{r,e}^{\beta_r}$ and $\eta=a_1+a_2q_{r,e}^{\beta_r}$ for the case $\gamma=\beta_c=1$.
\end{itemize}
The eigenvalue $-d\zeta q_{r,e}^{\zeta-1}$ is always negative, whereas
the sign of the eigenvalue $\lambda_{1}$ may
be positive or negative, depending on how large the supersaturation is
in comparison with the values $a_1\gamma$ and $a_2\beta_c$.
If the supersaturation $S$ is small enough and $\gamma=1$ or
$\beta_c=1$, the equilibrium point is stable. The stability behavior is
different in the remaining case $\gamma>1$ and $\beta_c>1$. For this
choice, the equilibrium point is always unstable.
Regarding our three cloud schemes, we find that the trivial equilibrium
point of the IFS scheme is  unstable, whereas the Wacker and
the COSMO scheme may admit a stable equilibrium if the supersaturation
is small enough.
This is a fundamental different behavior between the three schemes.

A numerical example of the coordinates of the trivial equilibrium point
is presented in table \ref{tab:qualitative_FP} (column ``trivial''),
where we assumed an environmental pressure \SI{1000}{\hecto\pascal},
environmental temperature \SI{273}{\kelvin}, supersaturation
\SI{0.1}{\percent} and $B=\num{e-3}$.
We indicated the coordinates of the equilibrium point with and without
evaporation, confirming the minor importance of this process. Note
that the trivial points of the different schemes are very close to
each other. Also note, that the non-scaled values are obtained by
multiplying the displayed values with
$q_{\mathrm{ref}}=\SI{e-4}{\kilogram\per\kilogram}$.

\begin{table}
  \centering
  \begin{tabular}{c|c|c|c|c}
    Scheme & \multicolumn{2}{c|}{Trivial} & \multicolumn{2}{c}{Non-Trivial} \\
     & $q_{c,e}$ & $q_{r,e}$ & $q_{c,e}$ & $q_{r,e}$ \\
    \hline 
    Wacker & $0$ & $0.258$ & $4.870$ & $6.533$ \\
    COSMO (without evaporation) & $0$ & $0.284$ & $7.939$ & $7.662$ \\
    COSMO & $0$ & $0.285$ & $7.943$ & $7.662$ \\
    IFS (without evaporation) & $0$ & $0.250$ & $3.045$ & $4.056$ \\
    IFS & $0$ & $0.259$ & $3.045$ & $4.056$ \\
  \end{tabular}
  \caption{Summary of the (nondimensional) equilibrium points, computed for an environmental pressure \SI{1000}{\hecto\pascal}, temperature \SI{273}{\kelvin} and supersaturation \SI{0.1}{\percent}. For the computations, we assumed a value $B=\num{e-3}$. The coordinates are rounded to three digits. The dimensional values can be obtained by multiplication with the factor \SI{e-4}{\kilogram\per\kilogram}.}
  \label{tab:qualitative_FP}
\end{table}

\subsection{The non-trivial Equilibrium Point}
\label{subsec:qualitative_nontrivial}

\begin{figure}
  \centering
  \subcaptionbox{Environmental temperature \SI{273}{\kelvin} and environmental pressure \SI{1000}{\hecto\pascal}.\label{fig:qualitative_non-trivial_FP-273}}[0.47\textwidth]{\includegraphics[width=0.47\textwidth]{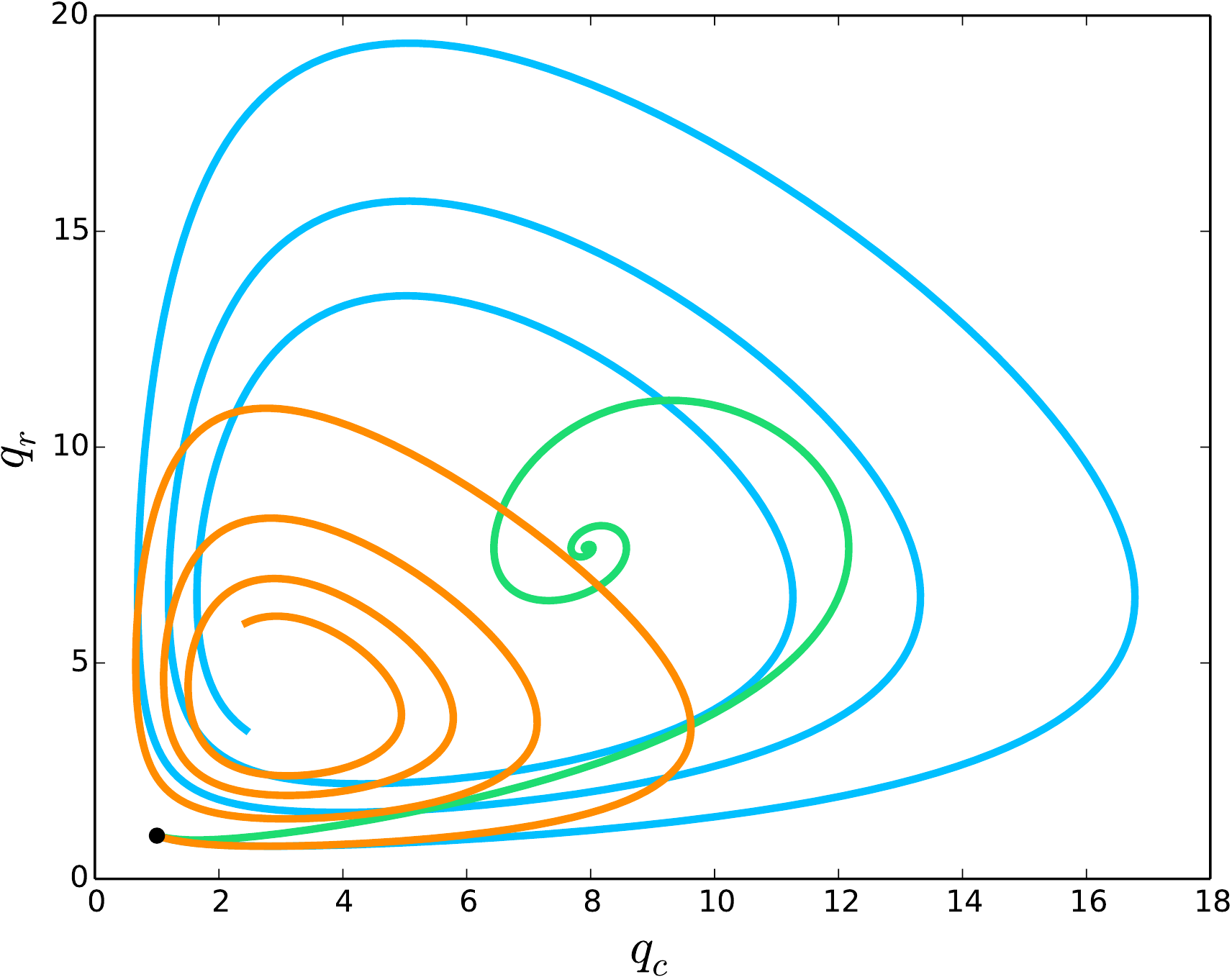}}
  \hspace{1em}
  \subcaptionbox{Environmental temperature \SI{250}{\kelvin} and environmental pressure \SI{800}{\hecto\pascal}.\label{fig:qualitative_non-trivial_FP-250}}[0.4\textwidth]{\includegraphics[width=0.47\textwidth]{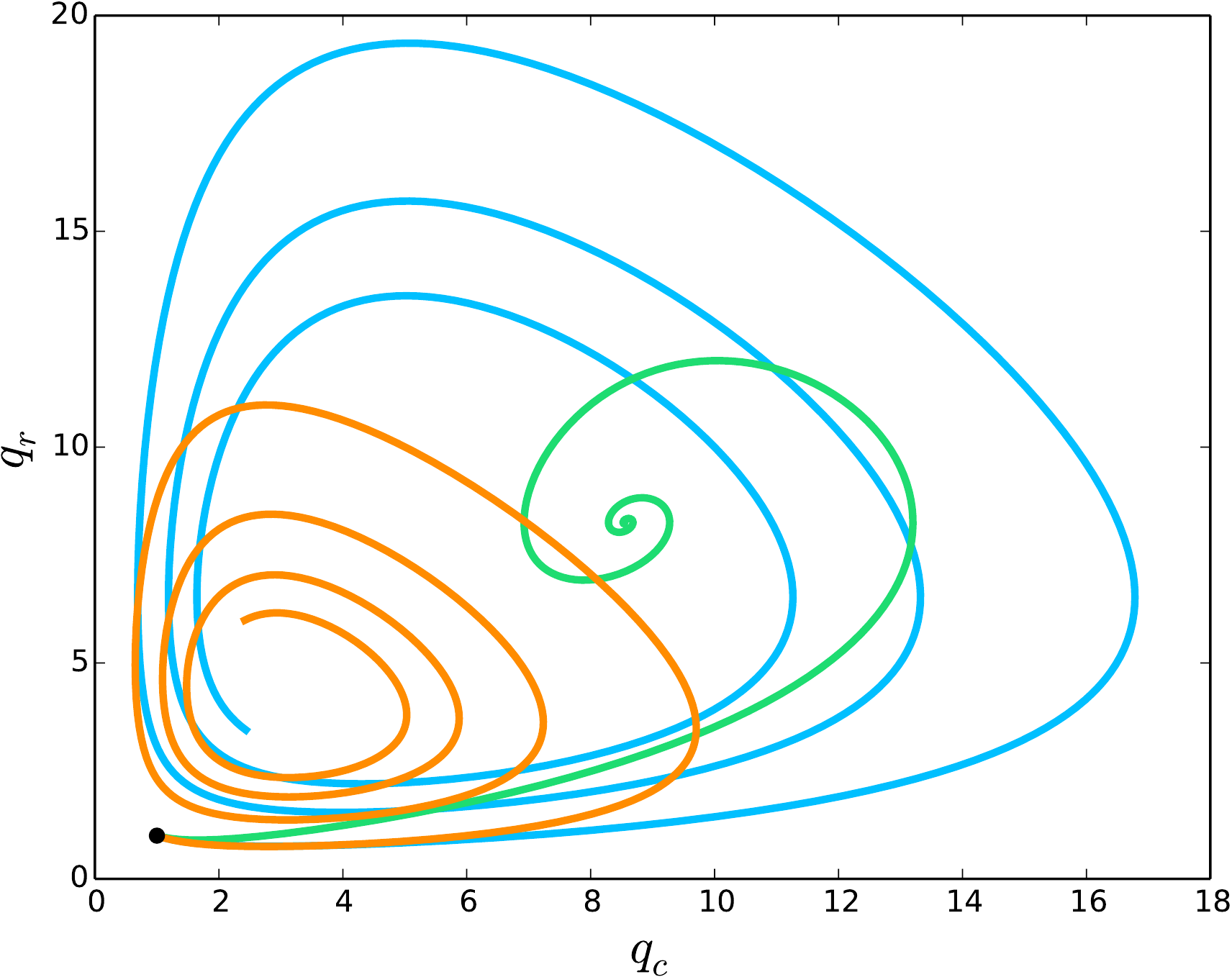}}
  \caption{Solution trajectories for all three cloud schemes with
    initial conditions $q_c\qty(0)=1$ and $q_r\qty(0)=1$, approaching their non-trivial equilibrium point for two different choices of the environmental conditions with $B=\SI{e-3}{\per\second}$. Wacker scheme: blue curve; COSMO scheme: green curve; IFS scheme: orange curve.}
  \label{fig:qualitative_non-trivial_FP}
\end{figure}

As already indicated, determining the equilibrium points
within the physical relevant range $0\leq q_c,\,q_r$ of the generic
cloud schemes is not possible in an analytical fashion. Therefore, we
restrict ourselves to the three cloud schemes Wacker, COSMO and IFS,
where we again neglect the evaporation process for the latter two. In
the following we always assume $q_{c,e}\neq0$; otherwise we recover
the trivial equilibrium point.

For the Wacker scheme, the equation $F(q_{c,e},\,q_{r,e})=0$ has only
one non-trivial root, given by
\begin{equation}
  \label{eq:qualitative_nontrivial_wacker}
  \begin{aligned}
    q_{c,e}=\frac{d}{a_2}-\frac{1}{cS}\qty[\frac{da_1}{a_2}+B]& &\text{and}& &q_{r,e}=\frac{cS-a_1}{a_2}.
  \end{aligned}
\end{equation}
The stability of this non-trivial equilibrium state depends on the magnitude
of $B$. As long as $B$ is not too small, the non-trivial equilibrium
point is stable \citep[see][]{wacker1992}.

We apply equation \eqref{eq:generic_generic_model-qc} with the
exponents of the COSMO scheme and arrive at 
\begin{equation}
  \label{eq:qualitative_nontrivial_cosmo-qr}
  q_{r,e}^{\frac{7}{8}}=\frac{cS-a_1}{a_2}.
\end{equation}
This equation is well-defined and admits a unique solution,
as long as $cS-a_1\geq0$, i.e. as long as
supersaturation is large enough. Assuming that this condition is met,
setting the right-hand side of
\eqref{eq:generic_generic_model-qr} to zero (neglecting $H$),
solving it for $q_c$ and substituting
\eqref{eq:qualitative_nontrivial_cosmo-qr}, we arrive at the unique solution
\begin{equation}
  \label{eq:qualitative_nontrivial_cosmo-qc}
  q_{c,e}=\frac{dq_{r,e}^{\frac{9}{8}}-B}{cS}.
\end{equation}

Considering the IFS scheme with its specific exponents, we set the
right-hand sides of
\eqref{eq:generic_generic_model-qc} 
and \eqref{eq:generic_generic_model-qr} to zero and add both
equations. This yields
\begin{equation}
  \label{eq:qualitative_nontrivial_ifs-qc}
  q_{c,e}=\frac{dq_{r,e}-B}{cS}.
\end{equation}
Substituting this relation into \eqref{eq:generic_generic_model-qc} results in the
non-linear equation
\begin{equation}
  \label{eq:qualitative_nontrivial_ifs-qr}
  0=cS-a_1\qty(\frac{dq_{r,e}-B}{cS})^{1.47}-a_2\qty(\frac{dq_{r,e}-B}{cS})^{0.15}q_{r,e}^{1.15}
\end{equation}
for $q_{r,e}$.

Apart from the computations above, we investigate the stability of the
equilibrium points numerically. Table
\ref{tab:qualitative_FP}
shows the (nondimensional) coordinates of the non-trivial equilibrium points
(column ``non-trivial'') for the same environmental conditions as above.
Figure \ref{fig:qualitative_non-trivial_FP} shows solution trajectories
for all three cloud schemes and two choices of the environmental conditions.
Observe that all trajectories spiral towards the non-trivial
equilibrium point, which indicates that these equilibrium points are stable.
Obviously, the equilibrium points of the three schemes are not identical.
This is a crucial observation:
simulating a warm cloud using the COSMO or the IFS scheme, leads to
different compositions of the mixing-ratios, as clouds are presumed to
be close to equilibrium. This feeds different values back to the other
parameterizations used in the forecast or climate model in which the
cloud schemes are incorporated and therefore can change the overall
model behavior. 

Inspecting the different trajectories in figure
\ref{fig:qualitative_non-trivial_FP}
reveals, that the number of spirals of the trajectory 
as well as the rate of convergence towards the respective equilibrium point
are different. We illustrate this behavior by computing the eigenvalues
$\lambda_1,\,\lambda_2$ of the derivative
$(DF)_{(q_{c,e},\,q_{r,e})}$
as well as the corresponding relaxation and
oscillation timescale, defined by
\begin{equation}
  \label{eq:qualitative_def_relax_oscillating_timescale}
  \begin{aligned}
    \tau_{\mathrm{relax}}=\frac{1}{\abs{\operatorname{Re}\qty(\lambda_1)}}& &\text{and}& &\tau_{\mathrm{osc}}=\frac{2\pi}{\abs{\operatorname{Im}\qty(\lambda_1)}}.
  \end{aligned}
\end{equation}
The results are summarized in table \ref{tab:qualitative_eigenvalues_etc}.
We observe that the relaxation timescale $\tau_{\mathrm{relax}}$ for
the COSMO scheme is significantly smaller than for the other two cloud schemes,
confirming the faster convergence of its trajectory towards the
equilibrium point in figure \ref{fig:qualitative_non-trivial_FP}
(green curve). Transferring this observation to the behavior of a warm cloud
as simulated by the COSMO scheme, we may expect that far more warm clouds
within this model are mostly in their equilibrium state compared to the
clouds, simulated by the other two cloud schemes.
In addition, figure \ref{fig:qualitative_non-trivial_FP} suggests that
the oscillation timescale $\tau_{\mathrm{osc}}$  of the trajectories
is smaller for the Wacker and the IFS scheme, compared to the COSMO scheme.
However, due to the smaller relaxation timescale, most windings of the green curve
are simply closer to the equilibrium point and are not visible in this
figure, although the oscillation timescale is indeed slightly larger for
the COSMO scheme (see table \ref{tab:qualitative_eigenvalues_etc}).

\begin{table}
  \centering
  \begin{tabular}{c|c|c|c}
    Scheme & $\lambda_1$  & $\tau_{\mathrm{relax}}$ & $\tau_{\mathrm{osc}}$ \\
    \hline 
    Wacker & $-\num{1.138e-4}+\num{4.272e-3}i$ & $\num{8787.346}$ & $\num{1470.821}$ \\
    COSMO & $-\num{1.173e-3}+\num{4.094e-3}i$ & $\num{852.56}$ & $\num{1534.583}$ \\
    IFS & $-\num{2.132e-4}+\num{4.953e-3}i$ & $\num{4690.822}$ & $\num{1268.59}$ \\
  \end{tabular}
  \caption{Eigenvalue $\lambda_1$ for the non-trivial equilibrium
    point for all three cloud schemes as well as the nondimensional relaxation and oscillation timescale, defined in \eqref{eq:qualitative_def_relax_oscillating_timescale}. The environmental conditions are the same as for table \ref{tab:qualitative_FP}. The values are rounded to three digits.}
  \label{tab:qualitative_eigenvalues_etc}
\end{table}

\section{Asymptotic Approach}
\label{sec:asymptotic}


In section
\ref{sec:qualitative}, we investigated the geometrical behavior of the
cloud schemes by computing their equilibrium points. As we pointed
out, the qualitative behavior is tightly connected to the long-time
behavior as $t\to\infty$. This raises the question about the
characteristic behavior of the system on shorter timescales. This is
a meaningful approach, since clouds will not experience the same
environmental conditions for very long (actually infinitely long)
times, rather the assumption of constant supersaturation is closely
related to the persistence of constant updrafts. Consequently,
the timescale of the updraft regime will determine meaningful timescales.
Thus we now consider the choice of several timescales for the analysis
and apply ideas of perturbation theory.



As indicated above, nondimensionalization of the governing
equations leads to certain constants
in the equations. The qualitative
behavior of the solution trajectories is connected to the relative
magnitudes of the constants. If only one small or large coefficient
were present, we could assume this parameter as $\epsilon$ or
$\epsilon^{-1}$ for a small $\epsilon$ and determine the behavior
of the solution trajectories by considering the limit $\epsilon\to0$.
However, our governing equation contains several coefficients.
Therefore, we choose a so-called Distinguished Limit, tying the
values of the constants for the different cloud schemes to a single
small parameter $\epsilon$
\citep[see, e.g.,][]{klein_etal2010}.
For our choice of the Distinguished Limit, we use the bare values of
the nondimensional coefficients \eqref{eq:nondimensionalization_coeff}
for the different cloud schemes as given in the appendix
\ref{sec:appendix_coeff} and figure
\ref{fig:specialized_condensation_rate} for the condensation rate. The
condensation rate spans roughly one order of magnitude, so we choose
\begin{equation}
  \label{eq:asymptotic_DL_c}
  c=\epsilon^{\mu}c^{\ast}
\end{equation}
with $\mu\in\{-1,\,0\}$ and $c^{\ast}=\order{1}$ as $\epsilon\to0$.
This choice is compatible with $\epsilon\sim\frac{1}{10}$ as in
\citet{hittmeir_klein2017}. For the sake of simplicity, in the sequel we only
consider the case $\mu=0$. In order to model different (constant)
supersaturations, we set
\begin{equation}
  \label{eq:asymptotic_DL_S}
  S=\epsilon^{\alpha}
\end{equation}
with $\alpha>0$
and a change in $\alpha$ corresponds to a change in supersaturation.
In the sequel, we always assume $1<\alpha$ if not indicated otherwise,
ensuring supersaturations
smaller than \SI{10}{\percent}.
The choice for the remaining coefficients is summarized in table
\ref{tab:asymptotic_DL_other_coeff}.
In order to change the timescale of the system 
\eqref{eq:generic_generic_model}, we additionally carry out a  time
transformation with new time variable
\begin{equation}
  \label{eq:asymptotic_time_transformation}
  \tau=\epsilon^{\omega}t.
\end{equation}

\begin{table}
  \centering
  \begin{tabular}{c|c|c|c|c|c}
    Scheme & $a_1$  & $a_2$ & $e_1$ & $e_2$ & $d$ \\
    \hline 
    Wacker & $a_1^{\ast}\epsilon^4$ & $a_2^{\ast}\epsilon^3$ & -- & -- & $d^{\ast}\epsilon^3$ \\ 
    COSMO & $a_1^{\ast}\epsilon^3$ & $a_2^{\ast}\epsilon^3$ & $e_1^{\ast}\epsilon^3$ & $e_2^{\ast}\epsilon^3$ & $d^{\ast}\epsilon^3$ \\ 
    IFS & $a_1^{\ast}\epsilon^7$ & $a_2^{\ast}\epsilon^3$ & $e_1^{\ast}\epsilon^7$ & $e_2^{\ast}\epsilon^2$ & $d^{\ast}\epsilon^3$ \\ 
  \end{tabular}
  \caption{Choice of the Distinguished Limit for the coefficients of the schemes with $a_1^{\ast},\,a_2^{\ast},\,e_1^{\ast},\,e_2^{\ast},\,d^{\ast}=\order{1}$ as $\epsilon\to0$. Note that evaporation is neglected in the Wacker scheme.}
  \label{tab:asymptotic_DL_other_coeff}
\end{table}

Applying the time transformation $\dv{q_c}{\tau}=\epsilon^{-\omega}\dv{q_c}{t}$
and analogously for $q_r$, as well as substituting the choices for the
Distinguished Limit, we arrive at the following equations: for
the Wacker scheme
\begin{subequations}
  \label{eq:asymptotic_wacker_with_DL}
  \begin{align}
    \dv{q_c}{\tau}&=\epsilon^{\mu+\alpha-\omega}c^{\ast}q_c-\epsilon^{4-\omega}a_1^{\ast}q_c-\epsilon^{3-\omega}a_2^{\ast}q_cq_r,\label{eq:asymptotic_wacker_with_DL-qc}\\
    \dv{q_r}{\tau}&=\epsilon^{4-\omega}a_1^{\ast}q_c+\epsilon^{3-\omega}a_2^{\ast}q_cq_r-\epsilon^{3-\omega}d^{\ast}q_r+\epsilon^{-\omega}B,\label{eq:asymptotic_wacker_with_DL-qr}
  \end{align}
\end{subequations}
for the COSMO scheme
\begin{subequations}
  \label{eq:asymptotic_COSMO_with_DL}
  \begin{align}
    \dv{q_c}{\tau}&=\epsilon^{\mu+\alpha-\omega}c^{\ast}q_c-\epsilon^{3-\omega}a_1^{\ast}q_c-\epsilon^{3-\omega}a_2^{\ast}q_cq_r^{\frac{7}{8}},\label{eq:asymptotic_COSMO_with_DL-qc}\\
    \dv{q_r}{\tau}&=\epsilon^{3-\omega}a_1^{\ast}q_c+\epsilon^{3-\omega}a_2^{\ast}q_cq_r^{\frac{7}{8}}+\epsilon^{3+\alpha-\omega}\qty(e_1^{\ast}q_r^{\frac{1}{2}}+e_2^{\ast}q_r^{\frac{11}{16}})-\epsilon^{3-\omega}d^{\ast}q_r^{\frac{9}{8}}+\epsilon^{-\omega}B,\label{eq:asymptotic_COSMO_with_DL-qr}
  \end{align}
\end{subequations}
and for the IFS scheme
\begin{subequations}
  \label{eq:asymptotic_IFS_with_DL}
  \begin{align}
    \dv{q_c}{\tau}&=\epsilon^{\mu+\alpha-\omega}c^{\ast}q_c-\epsilon^{7-\omega}a_1^{\ast}q_c^{2.47}-\epsilon^{3-\omega}a_2^{\ast}q_c^{1.15}q_r^{1.15},\label{eq:asymptotic_IFS_with_DL-qc}\\
    \dv{q_r}{\tau}&=\epsilon^{7-\omega}a_1^{\ast}q_c^{2.47}+\epsilon^{3-\omega}a_2^{\ast}q_c^{1.15}q_r^{1.15}+\epsilon^{7+\alpha-\omega}e_1^{\ast}q_r^{\frac{10}{9}}+\epsilon^{2+\alpha-\omega}e_2^{\ast}q_r^{\frac{127}{360}}\nonumber\\
    &\quad-\epsilon^{3-\omega}d^{\ast}q_r+\epsilon^{-\omega}B.\label{eq:asymptotic_IFS_with_DL-qr}
  \end{align}
\end{subequations}

As already indicated in the previous sections, we are now in a position to consider
different regimes by choosing appropriate values for the time transformation
exponent $\omega$ and the supersaturation exponent $\alpha$.
After the choice of all exponents, we consider a regular perturbation expansion
for the mixing-ratios
\begin{subequations}
  \label{asymptotics_expansion}
  \begin{align}
    q_c(\tau)&=q_c^{(0)}\qty(\tau)+\epsilon q_c^{(1)}\qty(\tau)+\order{\epsilon^2},\label{asymptotics_expansion-qc}\\
    q_r(\tau)&=q_r^{(0)}\qty(\tau)+\epsilon q_r^{(1)}\qty(\tau)+\order{\epsilon^2},\label{asymptotics_expansion-qr}
  \end{align}
\end{subequations}
substitute the expansions into the rescaled equations
\eqref{eq:asymptotic_wacker_with_DL},
\eqref{eq:asymptotic_COSMO_with_DL} and
\eqref{eq:asymptotic_IFS_with_DL} and collect the resulting 
reduced equations for the various orders of $\epsilon$.
An inconvenience arises, because the 
magnitude of the rain flux from above $B$ depends strongly on the actual
conditions. As a consequence, $B$ may be weak or strong and show up in any
order of the asymptotic expansion. This explains why we did not include
$B$ in the Distinguished Limit defined above. In the following, we
choose
\begin{equation}
  \label{eq:asymptotics_expansion_B}
  B=\epsilon^3\qty(B^{(0)}+\epsilon B^{(1)}+\order{\epsilon^2})
\end{equation}
as an expansion for $B$. 
This choice ensures that $B$ and $D$ are of the same asymptotic order
of magnitude and the terms may balance.
We emphasize that this is an assumption, one could also have
larger values for $B$. In this case, it would be necessary to use an
expansion as, e.g., 
$B=\epsilon^3\qty(\epsilon^{-1}B^{(-1)}+B^{(0)}+\epsilon
B^{(1)}+\order{\epsilon^2})$.
However, if $B \gg D$, the amount of water falling from above into the control
volume would be much larger than the amount of water falling out of the control
volume, resulting in an accumulation of water. On the other hand, if
$B \ll D$, the amount of water falling into the control
volume would be much smaller than the amount of water falling out and
the control volume would finally drain.



In the following section
\ref{subsec:reduced_eq}, we describe the behavior of the cloud schemes for different
regimes by inspecting the corresponding leading order reduced equations.

Another motivation to consider the reduced equation is given from a more
technical point of view. Suppose the cloud scheme is incorporated into a numerical
code and assume the numerical method would provide us with a sampling of the
exact solution, i.e. the numerical method would evaluate the exact solution.
In this case, we are given a sequence of discrete samplings
$\Gamma=\{(q_c\qty(n\Delta t),\,q_r(n\Delta t))\,\,|\,\,0\leq n\in\mathbb Z\}$ with the
timestep $\Delta t$. We may consider the numerical timestep as a timescale
for the governing equation and analyze the governing equation for this
particular timescale. From this we expect to get insight into the behavior
of the discrete samplings $\Gamma$ and consequently on the simulated
cloud.


\subsection{Derivation of the Reduced Equations}
\label{subsec:reduced_eq}


After having derived the nondimensional model equations
\eqref{eq:asymptotic_wacker_with_DL},
\eqref{eq:asymptotic_COSMO_with_DL},
\eqref{eq:asymptotic_IFS_with_DL}
for each cloud scheme
together with the corresponding Distinguished Limit, we can derive the
reduced equations for several regimes and timescales. We investigate
the behavior of the cloud schemes on the timescales
\begin{itemize}
\item $\omega=4$, representing a dimensional time $t'=t_{\mathrm{ref}}t=\epsilon^{-\omega}t_{ref}\tau=\epsilon^{-4}t_{\mathrm{ref}}\sim\SI{10000}{\second}$ for $\tau=1$,
\item $\omega=3$, representing a dimensional time \SI{1000}{\second},
\item $\omega=2$, representing a dimensional time \SI{100}{\second}, and
\item $\omega=1$, representing a dimensional time \SI{10}{\second}.
\end{itemize}
The long timescale of \SI{1000}{\second} is comparable to timesteps
within a climate model or the lifetime of a typical
Cumulus cloud \citep[e.g.,][]{book_rogers_yau1989}.
Therefore, an analysis on this timescale provides insight into the
behavior of the cloud schemes for climate model timesteps and the
simulated cloud dynamics during a Cumulus lifetime. As stated above,
we assume a constant supersaturation. Examples of
atmospheric phenomena that are able to maintain
a constant, but low, supersaturation for a long time is a warm conveyor belt
or vertical motions along fronts of a large scale pressure system.

The intermediate timescale of \SI{100}{\second} is comparable to the timestep
in numerical weather forecast models. On this timescale, we also expect only
low supersaturations, which may be maintained by persistent vertical motions.

Finally, the short timescale of
\SI{10}{\second} is more appropriate for Large Eddy Simulations or Cloud
Resolving models. On this timescale, supersaturations may attain larger values,
for example due to an updraft within a cloud core.

\subsubsection{Very Long Timescale \SI{10000}{\second}}
\label{subsubsec:reduced_eq_10000}

Considering a very long timescale \SI{10000}{\second}, by choosing
$\omega=4$, leads to
algebraic equations in leading order. For this timescale, we assume a
low supersaturation with $\alpha > 3$.
Applying an asymptotic expansion for $q_c$ and $q_r$,
we get the leading order equation for the Wacker scheme
\begin{subequations}
  \label{reduced_eq_10000_wacker_a4}
  \begin{align}
    0&=-a_2^{\ast}q_c^{(0)}q_r^{(0)},\label{reduced_eq_10000_wacker_a4-qc}\\
    0&=a_2^{\ast}q_c^{(0)}q_r^{(0)}-d^{\ast}q_r^{(0)} + B^{(0)}\label{reduced_eq_10000_wacker_a4-qr}
  \end{align}
\end{subequations}
for the COSMO scheme
\begin{subequations}
  \label{reduced_eq_10000_cosmo_a4}
  \begin{align}
    0&=-a_1^{\ast}q_c^{(0)}-a_2^{\ast}q_c^{(0)}\qty(q_r^{(0)})^{\frac{7}{8}},\label{reduced_eq_10000_cosmo_a4-qc}\\
    0&=a_1^{\ast}q_c^{(0)}+a_2^{\ast}q_c^{(0)}\qty(q_r^{(0)})^{\frac{7}{8}}-d^{\ast}\qty(q_r^{(0)})^{\frac{9}{8}}+
       B^{(0)}\label{reduced_eq_10000_cosmo_a4-qr}
  \end{align}
\end{subequations}
and for the IFS scheme
\begin{subequations}
  \label{reduced_eq_10000_ifs_a4}
  \begin{align}
    0&=-a_2^{\ast}\qty(q_c^{(0)}q_r^{(0)})^{1.15},\label{reduced_eq_10000_ifs_a4-qc}\\
    0&=a_2^{\ast}\qty(q_c^{(0)}q_r^{(0)})^{1.15}-d^{\ast}q_r^{(0)} + B^{(0)}.\label{reduced_eq_10000_ifs_a4-qr}
  \end{align}
\end{subequations}
The solution for all three cloud schemes is given by $q_c^{(0)} = 0$ and equation
\eqref{eq:trivial_fp}, i.e. on the time scale of \SI{10000}{\second}
the leading
order asymptotic solution coincides with the trivial equilibrium point,
representing a cloud-free scenario, where the incoming rain from above
falls through the air parcel.
For even longer
time scales and low supersaturations ($\alpha > 3$), we obtain the same
leading order equations.

\subsubsection{Long Timescale \SI{1000}{\second}}
\label{subsubsec:reduced_eq_1000}

We start with the long timescale by choosing $\omega=3$ and a supersaturation
of $S\sim\SI{0.1}{\percent}$, corresponding to $\alpha=3$. The resulting
leading order equations for the Wacker scheme are given by
\begin{subequations}
  \label{reduced_eq_1000_wacker}
  \begin{align}
    \dv{q_c^{(0)}}{\tau}&=c^{\ast}q_c^{(0)}-a_2^{\ast}q_c^{(0)}q_r^{(0)},\label{reduced_eq_1000_wacker-qc}\\
    \dv{q_r^{(0)}}{\tau}&=a_2^{\ast}q_c^{(0)}q_r^{(0)}-d^{\ast}q_r^{(0)}+B^{(0)},\label{reduced_eq_1000_wacker-qr}
  \end{align}
\end{subequations}
for the COSMO scheme,
\begin{subequations}
  \label{reduced_eq_1000_cosmo}
  \begin{align}
    \dv{q_c^{(0)}}{\tau}&=c^{\ast}q_c^{(0)}-a_1^{\ast}q_c^{(0)}-a_2^{\ast}q_c^{(0)}\qty(q_r^{(0)})^{\frac{7}{8}},\label{reduced_eq_1000_cosmo-qc}\\
    \dv{q_r^{(0)}}{\tau}&=a_1^{\ast}q_c^{(0)}+a_2^{\ast}q_c^{(0)}\qty(q_r^{(0)})^{\frac{7}{8}}-d^{\ast}\qty(q_r^{(0)})^{\frac{9}{8}}+B^{(0)},\label{reduced_eq_1000_cosmo-qr}
  \end{align}
\end{subequations}
and for the IFS scheme
\begin{subequations}
  \label{reduced_eq_1000_ifs}
  \begin{align}
    \dv{q_c^{(0)}}{\tau}&=c^{\ast}q_c^{(0)}-a_2^{\ast}\qty(q_c^{(0)}q_r^{(0)})^{1.15},\label{reduced_eq_1000_ifs-qc}\\
    \dv{q_r^{(0)}}{\tau}&=a_2^{\ast}\qty(q_c^{(0)}q_r^{(0)})^{1.15}-d^{\ast}q_r^{(0)}+B^{(0)}.\label{reduced_eq_1000_ifs-qr}
  \end{align}
\end{subequations}
It is remarkable that the Wacker and the IFS schemes are
essentially the same, since the exponent $1.15$ is comparable to $1$.
Regarding the accretion term, also the COSMO scheme is comparable to the others,
because the exponent $\frac{7}{8}$ is also comparable to $1$. However,
the COSMO scheme takes the autoconversion term in leading order into
account. All
schemes show a generalized predator-prey dynamics on this timescale and
supersaturation. For the Wacker scheme, the leading order system
\eqref{reduced_eq_1000_wacker} with $B^{(0)}=0$
is actually a Hamiltonian System and admits the
invariant
\begin{equation}
  \label{eq:reduced_eq_1000_wacker_invariant}
  I\qty(q_c^{(0)},\,q_r^{(0)})=a_2^{\ast}\qty(q_c^{(0)}+q_r^{(0)})-c^{\ast}\log(q_r^{(0)})-d^{\ast}\log(q_c^{(0)}),
\end{equation}
allowing periodic solutions \citep[see e.g.][]{book_verhulst1996}.
If $B^{(0)}\neq0$, the solutions converge towards the non-trivial
equilibrium point.

\begin{figure}
  \includegraphics[width=\textwidth]{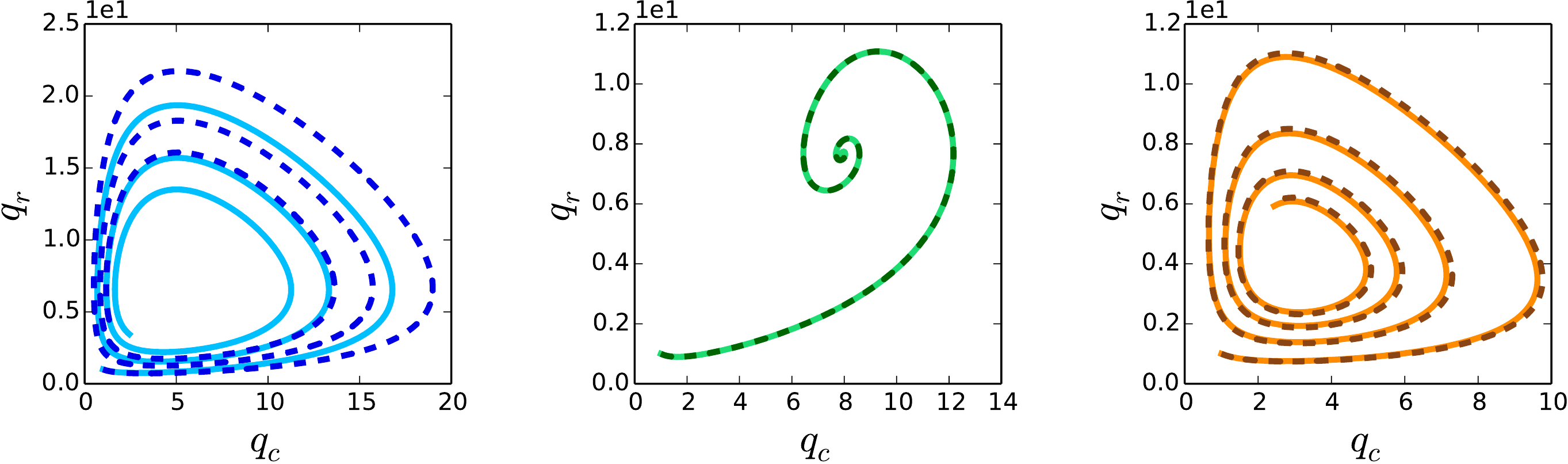}
  \caption{Numerical simulations of the full cloud schemes and the reduced
    equations on the timescale \SI{1000}{\second} with a constant
    supersaturation \SI{0.1}{\percent} and $B^{(0)}=1$ while
    $B^{(n)}=0$ for $n\geq2$; left panel: Wacker scheme; middle panel:
    COSMO scheme; right panel: IFS scheme. The solid line indicates
    the solution of the full cloud scheme and the dashed line the
    solution of the reduced equations.} 
  \label{fig:reduced_eq_1000_comparison}
\end{figure}

Figure
\ref{fig:reduced_eq_1000_comparison}
shows numerical simulations of the full schemes as well as the leading order
equations with $B^{(0)}=1$ and $B^{(n)}=0$ for $n\neq0$.
Examining the  figure shows that the reduced
equations agree very well with the full scheme and therefore
contain all dynamics on this timescale.

Considering a higher supersaturation with $\alpha<3$ yields the combination
\begin{equation}
  \label{eq:reduced_eq_1000_higher_S}
  \begin{aligned}
    q_c^{(0)}=0& &\text{and}& &\dv{q_r^{(0)}}{\tau}=-d^{\ast}\qty(q_r^{(0)})^{\zeta}+B^{(0)}
  \end{aligned}
\end{equation}
of an algebraic equation and a differential equation. Actually, one may
infer that $q_c^{(n)}=0$ for all orders $n$, such that the mixing-ratio
of cloud droplets
is exactly zero. According to \eqref{eq:reduced_eq_1000_higher_S},
rain water mass decreases and may only be compensated by the
rain flux from above. From \eqref{eq:reduced_eq_1000_higher_S} it is
evident, that the solution will approach the equilibrium point $(0,0)$
to leading order, as long as the source $B^{(0)}$ does not compensate this
convergence.
This behavior has no obvious physical interpretation, since for
large supersaturations, one would expect a new cloud to appear.
However, the choice $\alpha<3$ corresponds to supersaturations of
at least \SI{1}{\percent}, being a relatively high supersaturation
for this long timescale and may be considered unrealistic.


On the other hand, choosing a smaller supersaturation
corresponding to $\alpha\geq4$, yields the following reduced equations.
For the Wacker scheme we obtain
\begin{subequations}
  \label{reduced_eq_1000_wacker_a4}
  \begin{align}
    \dv{q_c^{(0)}}{\tau}&=-a_2^{\ast}q_c^{(0)}q_r^{(0)},\label{reduced_eq_1000_wacker_a4-qc}\\
    \dv{q_r^{(0)}}{\tau}&=a_2^{\ast}q_c^{(0)}q_r^{(0)}-d^{\ast}q_r^{(0)}+B^{(0)},\label{reduced_eq_1000_wacker_a4-qr}
  \end{align}
\end{subequations}
for the COSMO scheme,
\begin{subequations}
  \label{reduced_eq_1000_cosmo_a4}
  \begin{align}
    \dv{q_c^{(0)}}{\tau}&=-a_1^{\ast}q_c^{(0)}-a_2^{\ast}q_c^{(0)}\qty(q_r^{(0)})^{\frac{7}{8}},\label{reduced_eq_1000_cosmo_a4-qc}\\
    \dv{q_r^{(0)}}{\tau}&=a_1^{\ast}q_c^{(0)}+a_2^{\ast}q_c^{(0)}\qty(q_r^{(0)})^{\frac{7}{8}}-d^{\ast}\qty(q_r^{(0)})^{\frac{9}{8}}+B^{(0)},\label{reduced_eq_1000_cosmo_a4-qr}
  \end{align}
\end{subequations}
and for the IFS scheme,
\begin{subequations}
  \label{reduced_eq_1000_ifs_a4}
  \begin{align}
    \dv{q_c^{(0)}}{\tau}&=-a_2^{\ast}\qty(q_c^{(0)}q_r^{(0)})^{1.15},\label{reduced_eq_1000_ifs_a4-qc}\\
    \dv{q_r^{(0)}}{\tau}&=a_2^{\ast}\qty(q_c^{(0)}q_r^{(0)})^{1.15}-d^{\ast}q_r^{(0)}+B^{(0)}.\label{reduced_eq_1000_ifs_a4-qr}
  \end{align}
\end{subequations}
Using such a small supersaturation does not suffice to let the cloud droplets
grow by diffusion, resulting in a vanishing cloud to leading order.
It should be noted, that
in all three cloud schemes, the cloud does not vanish due to evaporation but
due to sedimentation of the rain drops, where the existing cloud
droplets are transformed into rain drops by collisional processes.
Again, note the similar exponents in the leading order equations, indicating
a similar behavior.

\subsubsection{Intermediate Timescale \SI{100}{\second}}
\label{subsubsec:reduced_eq_100}

Choosing $\omega=2$ yields the intermediate timescale, relevant for
numerical weather forecast models. When the reduced
equations are analyzed for different supersaturation (i.e. different
$\alpha$), one may realize
that the leading order equations are simply
\eqref{eq:reduced_eq_1000_higher_S} for $\alpha<2$.

Choosing $\alpha=2$ yields the leading order equations
\begin{equation}
  \label{eq:reduced_eq_100_herauslaufen_eq}
  \begin{aligned}
    \dv{q_c^{(0)}}{\tau}=c^{\ast}q_c^{(0)}& &\text{and}& &\dv{q_r^{(0)}}{\tau}=0,
  \end{aligned}
\end{equation}
since all processes are at least of order $\epsilon$, except condensation.
In this case, supersaturation is high enough to massively produce cloud
droplets in leading order, while autoconversion is too slow.
Note that in this case, the leading order solution for the cloud droplets is
unbounded, giving rise to secular terms and the asymptotic expansion breaks
down for longer nondimensional times $\tau$. This is a typical
situation in asymptotics however. When a solution is unbounded on the timescale
of $\SI{100}{\second}$, one should apply matched asymptotic solutions between
subsequent regimes from the hierarchy of timescales in order to construct
an approximation that is also valid within the subsequent regimes
\citep[see, e.g.,][for an introduction]{book_holmes2012}.
However, this is out of the scope of this paper and can be done in
future work. 



Decreasing supersaturation further on by choosing $2<\alpha$ yields the trivial
leading order equations
\begin{equation}
  \label{eq:reduced_eq_100_trivial_eq}
  \begin{aligned}
    \dv{q_c^{(0)}}{\tau}=0& &\text{and}& &\dv{q_r^{(0)}}{\tau}=0
  \end{aligned}
\end{equation}
whose solutions are constant in time and admit the initial conditions.
When first and higher order
corrections are constructed, these corrections are polynomials in
$\tau$ and again introduce secular behavior, indicating the need
for a matched asymptotic solution.



\subsubsection{Short Timescale \SI{10}{\second}}
\label{subsubsec:reduced_eq_10}

The choice $\omega=1$ selects a short timescale typical for Large Eddy
Simulations or Cloud Resolving models. As before, the leading order equation
\eqref{eq:reduced_eq_100_herauslaufen_eq} is found for $\alpha=1$,
representing the very large supersaturation of \SI{10}{\percent}.

Assuming a higher supersaturation yields always the leading order equations
\eqref{eq:reduced_eq_100_trivial_eq} 
such that the leading order solution reproduces the initial conditions.
If the supersaturation is chosen as \SI{1}{\percent}, condensation is
the dominant process in first order and $q_c^{(1)}$ increases linearly.
If the supersaturation is smaller than \SI{1}{\percent}, all
dynamics takes place in order $\order{\epsilon^2}$.
Note that on this timescale, we always encounter secular terms limiting 
the validity of the asymptotic approach to a relatively short (rescaled)
time interval. The resolution of this problem is, again, given by
constructing a matched asymptotic solution.


\subsection{Discussion}
\label{subsec:discussion}

In section \ref{subsec:qualitative_trivial}, we computed the trivial
equilibrium point analytically by neglecting the
evaporation. Examining the nondimensional 
equations
\eqref{eq:asymptotic_COSMO_with_DL},
\eqref{eq:asymptotic_IFS_with_DL}
for the COSMO and the IFS schemes reveals, that evaporation only gives a
higher order contribution, and does indeed not enter the leading order
equations above. This justifies neglecting the evaporation process
in hindsight.

Inspecting the equations
\eqref{eq:asymptotic_wacker_with_DL},
\eqref{eq:asymptotic_COSMO_with_DL} and
\eqref{eq:asymptotic_IFS_with_DL},
we realize that the importance of
the autoconversion process depends strongly on the cloud scheme.
Only the COSMO scheme takes account of autoconversion 
in leading order, which may lead to a faster
occurrence of rain compared to the other
models.
We remark, that some climate models use the same autoconversion rate
as the IFS scheme, following \citet{khairoutdinov_kogan2000}, but
change the actual rate by multiplication with a constant factor.
This corresponds to an artificial shift of the autoconversion process
into higher orders and is done to restore the radiation balance of the climate
model.

\begin{figure}
  \subcaptionbox{Regime $\mu=0$\label{fig:discussion_summary_mu_0}}[0.47\textwidth]{\includegraphics[width=0.47\textwidth]{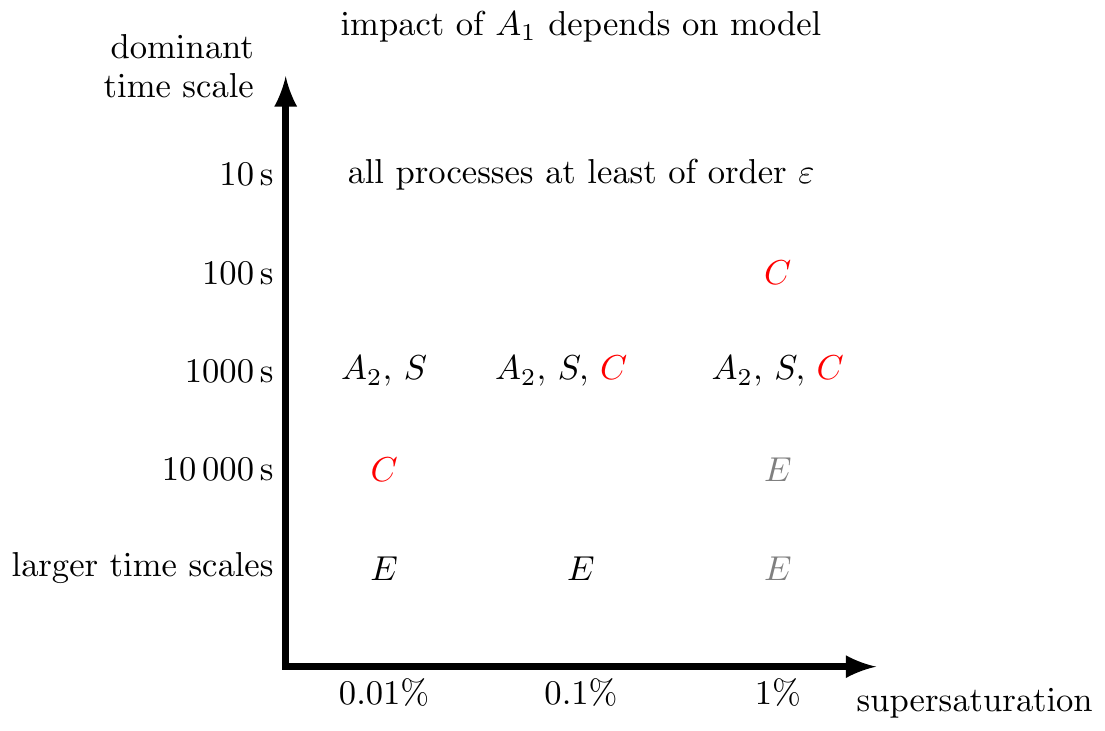}}
  \hspace{1em}
  \subcaptionbox{Regime $\mu=-1$\label{fig:discussion_summary_mu_1}}[0.47\textwidth]{\includegraphics[width=0.47\textwidth]{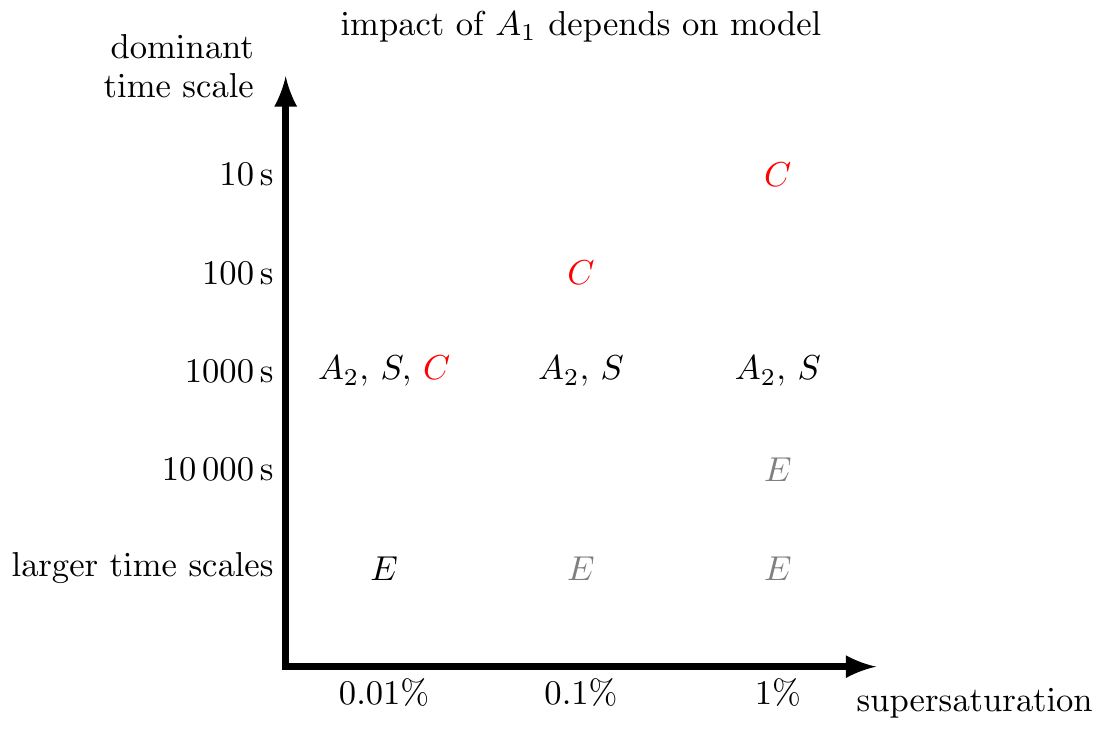}}
  \caption{Summary of the occurrence of the individual cloud processes in the leading order equations for a given timescale and supersaturation for both possible regimes $\mu=0$ (left panel) and $\mu=-1$ (right panel) of the condensation process. Regime $\mu=0$ is discussed in our study. Note that condensation and evaporation depend on the supersaturation.}
  \label{fig:discussion_summary}\eqref{eq:asymptotic_COSMO_with_DL},
\end{figure}

Figure \ref{fig:discussion_summary_mu_0} summarizes the occurrence
of the individual processes in the leading order equations,
depending on the timescale and supersaturation.
Since condensation and evaporation depend on the supersaturation,
these processes change their dominant timescale as the supersaturation
is altered, in particular the processes become faster for increasing
supersaturations.
As discussed above, the impact of the autoconversion depends strongly
on the cloud scheme and is therefore not shown in this figure.
Figure \ref{fig:discussion_summary_mu_1} illustrates the consequences
of choosing the regime $\mu=-1$ for the condensation, see equation
\eqref{eq:asymptotic_DL_c} and figure
\ref{fig:specialized_condensation_rate}.

As already pointed out, the chosen timescales correspond to typical timesteps
of atmospheric models, i.e. a Large Eddy Simulation model employs timesteps
of order \SI{1}{\second} to \SI{10}{\second}, a weather forecast model uses
timesteps of order \SI{100}{\second}
whereas a climate model uses timesteps of order \SI{1000}{\second} or larger.
By constructing the reduced equations of the cloud schemes, we analyzed the
inherent behavior of the cloud schemes on those timescales.
This helps in interpreting the interactions of the cloud scheme with
other parameterizations of the numerical model. For example, if the
numerical model employs a timestep of \SI{10}{\second}, we know from
section
\ref{subsubsec:reduced_eq_10}
that the cloud scheme essentially reproduces the given initial values.
Consequently, the other parameterizations of the numerical model will get
essentially the initial values as feedback from the cloud scheme.
In contrast, if the numerical model employs larger timesteps and the
reduced equations show dynamical behavior on the corresponding timescale, an accurate
approximation of the ODE, representing the cloud scheme, is needed.
This point of view might also be helpful in the
implementation of atmospheric models, since one can take the knowledge about the
behavior of the cloud schemes on certain timescales into account.
This information can be passed in a meaningful way to other parameterizations
in the models.

\section{Conclusions}
\label{sec:conclusion}

In this study, we presented a generic one moment cloud scheme for a warm cloud
in the spirit of
the classical Kessler schemes. The cloud scheme utilized in \citet{wacker1992}
as well as the cloud schemes implemented in the operational models COSMO and
IFS fit into this framework.
For our study, we used the cloud schemes essentially as they are documented
in \citet{wacker1992} or the model documentations
\citep{cosmo_doc_physical_parameterization,ifs_doc_physical_parameterization}.
We only changed the representation of the condensation process. In the study by
\citet{wacker1992}, an explicit parameterization is used, whereas the
operational schemes bypass an explicit formulation
by using saturation adjustment. In our study, we use an explicit
parameterization of condensation which is derived physically and employ this
formulation in all three specialized cloud schemes. Representing the
condensation process by the same parameterization in the three cloud schemes
re-establishes their comparability.

Moreover, we analyzed the qualitative behavior of the three cloud schemes by computing
equilibrium states as well as analyzing the stability of the equilibrium
states. Apart from the trivial equilibrium state, we found another
equilibrium state.
The trivial equilibrium state corresponds to a cloud-free case and is always
unstable for the IFS scheme. Considering the Wacker and the COSMO schemes, this
equilibrium may become stable for small supersaturations. As outlined in
table
\ref{tab:qualitative_FP},
the trivial equilibrium states are comparable but the
non-trivial equilibrium states cannot be regarded as very similar.
This implies an inherently different behavior
of the simulated clouds when we consider a long simulation time or
 the initial values are near a non-trivial equilibrium state, because
these equilibria are attractive in many cases.

Strictly speaking, an analysis of equilibrium states involves considering the
limiting behavior as $t\to\infty$. Therefore, we used asymptotics to infer
the characteristic behavior of the cloud schemes on shorter timescales.
The analysis shows that a timescale of \SI{10000}{\second} is enough
for each cloud scheme to arrive in a possible equilibrium state to leading
order.

On timescales comparable to \SI{1000}{\second} and a low supersaturation
\SI{0.1}{\percent}, all cloud schemes show nontrivial dynamics, similar to
a (forced) predator-prey system. Already in \citet{wacker1992}, predator-prey
dynamics can be found for a cloud scheme. The exponents of the
accretion term
are different in the three cloud schemes but are all comparable to $1$.
This indicates similar dynamical behavior, although the limiting equilibrium
states differ. 

For smaller supersaturations, all leading order equations for the
mixing ratio of cloud droplets only contain 
sinks. Physically speaking, we get a shrinking cloud.
In the considered case, the cloud does not vanish due to evaporation but due to
the conversion of all cloud droplets into rain drops and subsequent
sedimentation of the rain drops.
From a geometrical point of view, both equilibrium points change their
stability behavior such that the non-trivial equilibrium point becomes
unstable.

On the shorter timescales for high supersaturations,
we encounter a sharp increase of the cloud
droplets. This behavior is consistent with the underlying physics.
For smaller supersaturations,
the dynamics becomes trivial and the leading order simply reproduces
the initial conditions; all dynamics is deferred to higher orders.

With this study, we do not intend to rate the considered cloud schemes, but instead
establish their characteristic behavior. Knowing the characteristics
of the warm cloud schemes on different timescales and humidity regimes
helps in interpreting the outcome of the full operational model
with respect to the simulation of warm clouds. However, since
the interactions
between different parameterizations in a forecast or climate model
are very complex, the impact of differences within the warm cloud
schemes are quite unclear, but may be significant. A prominent example
of a complex change in the behavior may be observed by coupling a
nonlinear ODE, describing chemical reactions,
to a PDE by introducing diffusion terms
\citep[an example may be found in][]{barrio_etal1999}:
a stable equilibrium state of the ODE can be destabilized and eventually
form new spatial or temporal patterns
\citep[Turing instabilities, see, e.g.][]{cross_hohenberg1993,turing1952}.
In this sense, when the cloud schemes are coupled to the Navier-Stokes equations
for moist atmospheric flow, even the stable equilibrium states found in this
study may destabilize and give rise to pattern formation.


\section{Acknowledgement}

  We thank Rupert Klein and an anonymous referee for their helpful comments
  on our manuscript.
  Juliane Rosemeier acknowledges support of the Transregional Collaborative
  Research Center SFB/TRR 165 ``Waves to Weather'', funded by the
  ``Deutsche Forschungsgemeinschaft'' (DFG), within the subproject
  ``Structure Formation on Cloud Scale and Impact on Larger Scales''.
  Manuel Baumgartner acknowledges support of the ``Deutsche Forschungsgemeinschaft''
  (DFG) within the project ``Enabling Performance Engineering in Hesse and
  Rhineland-Palatinate'' (grant number 320898076).


\appendix

\section{Derivation of the Condensation Term}
\label{sec:appendix_condensation}

The equation, describing the change of the mass of a single cloud droplet due
to the diffusion of water vapor, is given by
\citep{book_rogers_yau1989}
\begin{equation}
  \label{eq:appendix_condensation_equation}
  \dv{m'}{t'}=\frac{4\pi r'S}{\qty(\frac{L'}{R_v'T'}-1)\frac{L'}{k'T'}+\frac{R_v'T'}{\alpha_d D'e_{\mathrm{sat}}(T')}}\eqqcolon A'\cdot\qty(m')^{\frac{1}{3}}S,
\end{equation}
where
$m'$ is the mass of the cloud droplet,
$r'$ the radius of the cloud droplet,
$L'$ the latent heat of vaporization,
$R_v'$ the individual gas constant for water vapor,
$T'$ the environmental temperature,
$k'$ the thermal conductivity of dry air,
$D'$ the diffusivity of dry air,
$\alpha_d$ the mass accommodation coefficient and
$e_{\mathrm{sat}}'$ the saturation vapor pressure of liquid water.

We describe an ensemble of water droplets by a size (or mass)
distribution with density $f'(m')$, which is normalized by the number
concentration $n_c'$, i.e. the number concentration is given by the
zeroth moment of the distribution. The mass mixing ratio can be
expressed by the first moment of the mass distribution,
i.e. $q_c'=\int_{\mathbb R}m'f'(t',m')\dd{m'}$.
The total time derivative of $q_c'$ is then given by 
\begin{equation}
  \label{eq:appendix_condensation_equation_derivation}
  \begin{aligned}
    \dv{q_c'}{t'}\qty(t')&=\dv{t'}\int\limits_0^{\infty}m'f'(t',m')\dd{m'}\\
    &=\int\limits_0^{\infty}f'(t',m')\dv{m'}{t'}\dd{m'}=A'S\int\limits_0^{\infty}f'(t',m')\qty(m')^{\frac{1}{3}}\dd{m'}
  \end{aligned}
\end{equation}
For this derivation, we applied the continuity equation for the
size distribution $f'$ in the phase-space,
i.e. $\pdv{f'}{t'}\qty(t',m')+\pdv{m'}(\dv{m'}{t'}f')=0$ in case of
no particle formation, partial integration as well as $f'(t',m')=0$
for $m'\leq0$ and the assumption that $f'$ decays fast \citep[see
also, e.g.,][]{spichtinger_gierens2009a}.

For all typical mass distributions, usually used in cloud physics, we
obtain analytical expressions for the general moments of the type
\begin{equation}
  \label{eq:appendix_general_moments}
  \begin{aligned}
    \int\limits_0^{\infty}f'(t',m')\qty(m')^r\dd{m'}=n_c{m_c'}^r\cdot c(f),& &r\in\mathbb{R}
\end{aligned}
\end{equation}
with the mean mass ${m_c'}=\frac{q_c'}{n_c'}$ and a correction
factor $c(f)$ depending on the type of the mass distribution
\citep[see, e.g.,][for generalized Gamma or lognormal
distributions]{seifert_beheng2006a,spichtinger_gierens2009a}.
Thus, the condensation rate for $q_c'$ can be described as
\begin{equation}
  \label{eq:appendix_condensation_rate_general}
  \dv{q_c'}{t'}\qty(t')=A'S n_c'{m_c'}^{\frac{1}{3}}\cdot c(f).
\end{equation}
Since the correction factor is usually of order $O(1)$, we
approximately set $c(f)\approx 1$ and thus 
\begin{equation}
  \label{eq:appendix_condensation_rate_approx}
  \dv{q_c'}{t'}\qty(t')=A'S n_c'{m_c'}^{\frac{1}{3}}.
\end{equation}
As in the derivation by \citet{wacker1992} we assume that during
condensation the mean size of droplets do not change
drastically. Thus, as a first approximation, we can assume
${m_c'}$ as a constant. Using the relation for the mean mass
$q_c'=n_c'{m_c'}$ we obtain the final description of the
condensation rate:
\begin{equation}
  \label{eq:appendix_condensation_equation_final}
  \dv{q_c'}{t'}\qty(t')=
  A'S\cdot\frac{q_c'}{{m_c'}}{m_c'}^\frac{1}{3}=
  \underbrace{
   A' {m_c'}^{-\frac{2}{3}}
  }_{\eqqcolon c'}Sq_c'.
\end{equation}
Note that a similar linear relation is used in the study by
\citet{klein_majda2006}, using asymptotic methods for convective clouds.

\section{Coefficients of the Cloud Schemes}
\label{sec:appendix_coeff}

In this appendix, we collect the values of the constant coefficients and
exponents and illustrate the dependency of the non-constant coefficients
for all three cloud schemes on the environmental conditions.
Table \ref{tab:appendix_coeff} collects all constant
coefficients and the exponents for all three cloud schemes. 
Note that for the Wacker scheme, all coefficients are constant, except the
condensation rate, which is equal among all three cloud schemes.

\begin{table}
  \centering
  \begin{tabular}{c|c|c|c|c|c|c}
    Coefficient & $a_1$ & $a_2$ & $e_1$ & $e_2$ & $d$  \\
    \hline
    Wacker & $\num{e-4}$ & $\num{7.5e-4}$ & $0$ & $0$ & $\num{3.88e-3}$ & \\
    COSMO & $\num{e-3}$ & X & X & X & X & \\
    IFS & $\num{9.83e-8}$ & $\num{8.45e-4}$ & X & X & $\num{4e-3}$ &\\
    \hline
    Exponent & $\gamma$ & $\beta_c$ & $\beta_r$ & $\delta_1$ & $\delta_2$ & $\zeta$ \\
    \hline
    Wacker &  $1$ & $1$ & $1$ & $1$ & $1$ & $1$ \\
    COSMO &  $1$ & $1$ & $\frac{7}{8}$ & $\frac{1}{2}$ & $\frac{11}{16}$ & $\frac{9}{8}$ \\
    IFS & $2.47$ & $1.15$ & $1.15$ & $\frac{10}{9}$ & $\frac{127}{360}$ & $1$ \\
  \end{tabular}
  \caption{Values of the constant coefficients and the exponents for all three cloud schemes. A non-constant coefficient is indicated by X.}
  \label{tab:appendix_coeff}
\end{table}

\subsection{COSMO}
\label{subsec:appendix_coeff_cosmo}


The non-constant coefficients of the COSMO scheme are found in
\citet{cosmo_doc_physical_parameterization} in their equations
(5.46) for $a_2$, (5.47) for $e_1$ and $e_2$, (5.41) for $d$.

\begin{figure}
  \centering
  \includegraphics[width=0.8\textwidth]{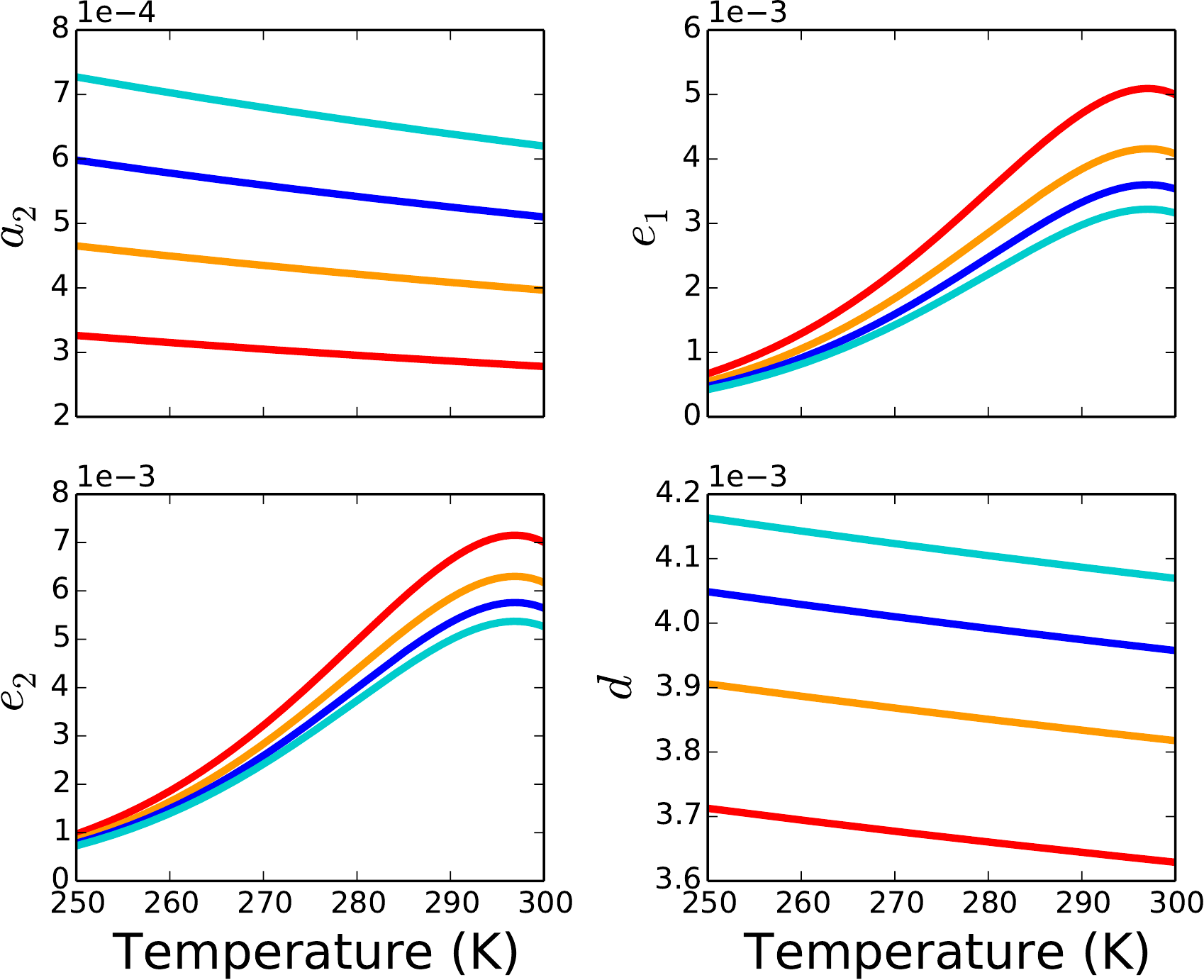}
  \caption{Nondimensional non-constant coefficients
    $a_2,\,e_1,\,e_2,\,d$ for the COSMO scheme as functions of temperature for various pressures: red curve: \SI{400}{\hecto\pascal}; orange curve: \SI{600}{\hecto\pascal}; blue curve: \SI{800}{\hecto\pascal}; cyan curve: \SI{1000}{\hecto\pascal}.}
  \label{fig:appendix_coeff_cosmo}
\end{figure}

Figure \ref{fig:appendix_coeff_cosmo} shows these coefficients as functions
of temperature for various pressures.
It is easily seen, that the coefficients are essentially constant
regarding their order of magnitude, although they depend weakly on temperature
and pressure.

\subsection{IFS}
\label{subsec:appendix_coeff_ifs}

The non-constant coefficients of the IFS scheme are found in
\citet{ifs_doc_physical_parameterization}
in their equation (7.75)
for $e_1$ and $e_2$.

\begin{figure}
  \centering
  \includegraphics[width=0.8\textwidth]{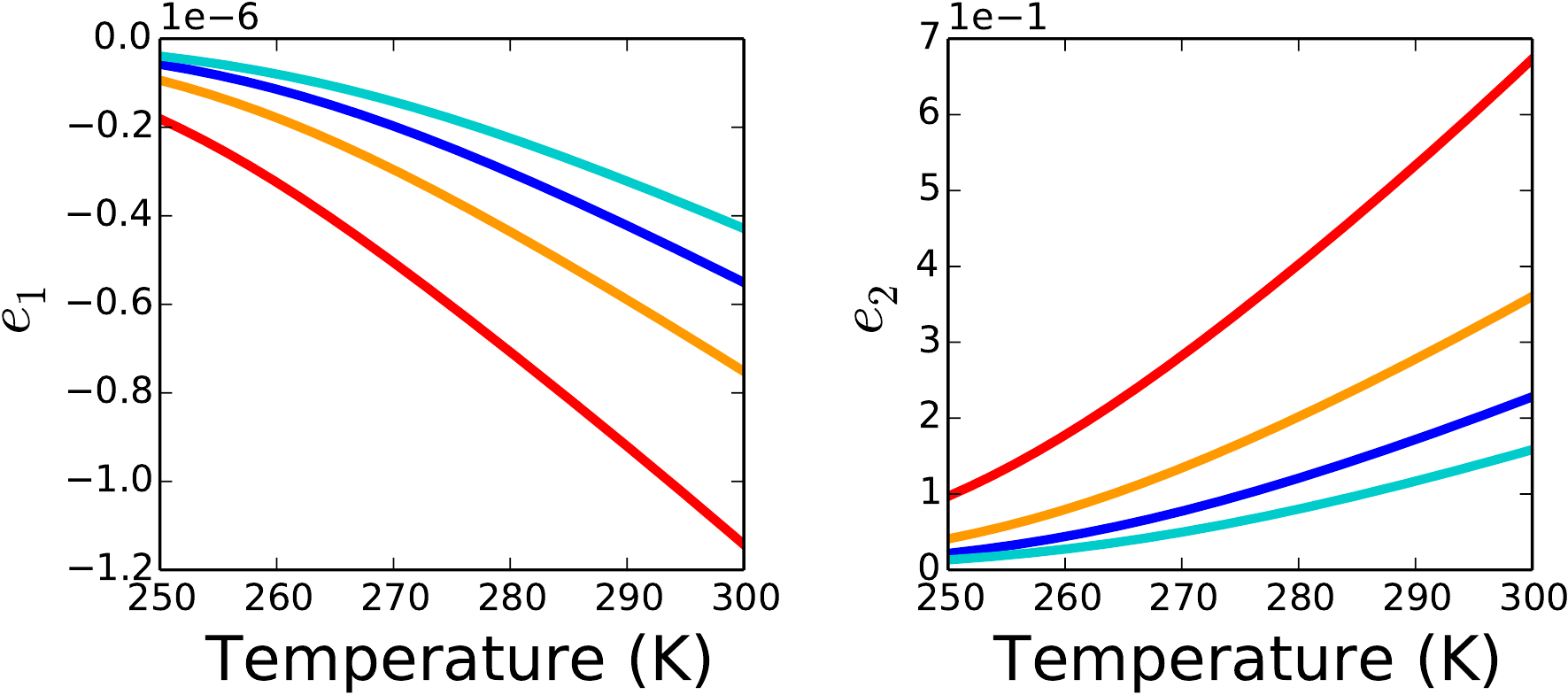}
  \caption{Nondimensional non-constant coefficients $e_1,\,e_2$ for
    the IFS scheme as functions of temperature for various pressures: red curve: \SI{400}{\hecto\pascal}; orange curve: \SI{600}{\hecto\pascal}; blue curve: \SI{800}{\hecto\pascal}; cyan curve: \SI{1000}{\hecto\pascal}.}
  \label{fig:appendix_coeff_ifs}
\end{figure}

Figure \ref{fig:appendix_coeff_ifs} shows the non-constant nondimensional coefficients
for the IFS scheme as functions of temperature for various pressures.
Also in this case, the values of the non-constant coefficients may be
considered as roughly
constant for our asymptotic analysis.
Note that the formulation of the saturation vapor pressure in the description of
the IFS scheme differs slightly from ours, since we used the accurate
formulation from \citet{murphy_koop2005}.

\printbibliography

\end{document}